\documentclass[12pt,epsfig]{article}
\usepackage{amsfonts}
\usepackage{graphicx}
\usepackage{amsmath}

\input epsf.sty

\def\BC{\bb C}
\def\_\BC{\bbi C}

\input{TCILATEX.TEX}

\begin{document}
\begin{titlepage}
\setcounter{page}{1}
\renewcommand{\thefootnote}{\fnsymbol{footnote}}

\begin{flushright}
GNPHE/0402\\
CDUTPG 04-01\\
UMDEPP 05-018\\
%hep-th/yymmxxx\\
\end{flushright}

%\vspace{2mm}
\begin{center}
{\Large\bf{Supersymmetric Embedding of the Quantum\\ Hall Matrix
Model}}

\vspace{6mm}

{\small\bf S. James Gates Jr$^{1}$\footnote{gatess@wam.umd.edu},
Ahmed Jellal$^{2}$\footnote{jellal@ucd.ac.ma}, EL Hassan
Saidi$^{3}$\footnote{ H-saidi@fsr.ac.ma} and Michael
Schreiber$^{4}$ \footnote{schreiber@physik.tu-chemnitz.de}}
\vspace{4mm}

$^{1}${\small{\em Center for String and Particle Theory, Physics
Department, University of Maryland,}}\\
{\small\em College Park, MD 20742-4111 USA}

$^{2}${\small \em Theoretical Physics Group, %Physics Department,
Faculty of Sciences, Chouaib Doukkali University},\\
{\small\em Ibn Ma\^achou Road, P.O. Box 20, 24000 El Jadida,
Morocco}

 \emph{}$^{3}${\small \em Groupement National de Physique des
Hautes Energies (GOPHER), Facult\'{e} des Sciences, }\\
{\small\em Av Ibn Batouta, P.O. Box 1014, Rabat, Morocco}

 $^{4}${\small \em Institut f\"ur Physik, Technische Universit\"at
 Chemnitz, D-09107 Chemnitz, Germany}\\

\end{center}

\vspace{5mm}
\begin{abstract}
We develop a supersymmetric extension of the
Susskind-Polychronakos matrix theory for the quantum Hall fluids.
This is done by considering a system combining two sets of
different particles and using both a component field method as
well as world line superfields. Our construction yields a class of
models for fractional quantum Hall systems with two phases U and D
involving, respectively $N_1$ bosons and $N_2$ fermions. We build
the corresponding supersymmetric matrix action, derive and solve
the supersymmetric generalization of the Susskind-Polychronakos
constraint equations. We show that the general $U(N)$ gauge
invariant solution for the ground state involves two
configurations parameterized by the bosonic contribution $k_{1}$
(integer) and in addition a new degree of freedom $k_{2}$, which
is restricted to $0$ and $1$. We study in detail the two
particular values of $k_{2}$ and show that the classical
(Susskind) filling factor $\nu $ receives no quantum correction.
We conclude that the Polychronakos effect is exactly compensated
by the opposite fermionic contributions.
\end{abstract}

\bigskip
\begin{quote}
\end{quote}

\end{titlepage}
\newpage

\tableofcontents

\section{Introduction}

In a work by Susskind~\cite{susskind} it was asserted that the
non-commutative Chern-Simons (NCCS) gauge theory in $(2+1)$
dimensional space is equivalent to Laughlin theory~\cite{laughlin}
at filling factor $\nu =\frac{1}{k}$, with $k$ a positive (odd)
integer. This formulation leads to a matrix model similar to that
describing $D0$-branes in string theory and has opened a revival
interest in the study of the fractional quantum Hall
effect~\cite{tsui} in terms of NCCS theory. In particular a
regularized version has been proposed by Polychronakos~\cite{
polychronakos} where a quantum correction to the Susskind filling
factor and the corresponding ground state has been
constructed~\cite{hellerman1}. Other investigations about the
relation between NCCS and Laughlin fluids can be found in recent
works~\cite{sakita,kobashi,park}. The Susskind model and its
regularized version has been extended to quantum Hall (QH) states
that are not of the Laughlin type in a multi-component
Chern-Simons approach~\cite{saidi1} and another based on the
Haldane hierarchy~\cite{haldane} has been developed~\cite{
jellal2,jellal3}.  Also a matrix model for bilayered QH
systems~\cite{jellal4} has been considered as well as an isotropic
QH one~\cite{jellal5}.

In this paper we propose a supersymmetric matrix model (SMM) for
quantum Hall fluids and thus generalized the original
Susskind--Polychronakos (SP) theory for the Laughlin states.  It
can be used to describe a class of fractional QH systems resulting
from two sets U and D of $N_1$ bosons and $N_2$ fermions. In our
proposal, the states of these sets are imagined as two states of a
world line supersymmetric representation with $N_{1}=N_{e}$ and
$N_{2}=N_{\widetilde{e}}$ where $\widetilde{e}$ is the
superpartner of the particle $e$. Our supersymmetric extension
(SE) of the SP model gives new solutions for Laughlin states with
the total filling factor $\frac{1}{k}$. More precisely, we
consider a QH system of $N$ bosons and fermions $ (N=N_{1}+N_{2})$
with $N_{\phi }=kN$ quantum fluxes $(N_{\phi
}=k_{1}N_{1}+k_{2}N_{2})$ and investigate its basic features.

To achieve this proposal, we develop a SMM for a fractional QH
system and show how this can describe U and D together. We write
the SMM action and derive the supersymmetric constraint equations.
These will be solved to obtain wave functions for the lowest
Landau level and the corresponding filling factors. Throughout
this study, we obtain new results which can be interpreted in
terms of supersymmmetric representation theory. In particular, we
find two Polychronakos effects coming with opposite signs and
cancel each other exactly in agreement with Bose-Fermi
cancellation.

Before doing this task, we develop and discuss another interesting
result. This concerns a general relation governing filling factor
and energy of a given QH system. It allows us to study the
$\frac{1}{N}$\ corrections in fractional QH systems and therefore
give some information on the behavior of the collective motions of
the interacting QH particles. This has another consequence, mainly
it can be used to derive an immediate description of a QH system
combining bosons and fermions.

The presentation of this paper is as follows. In section 2, we
revisit the main steps of the SP construction. A link between
filling factor $\nu $ and the energy will be given as well as a
$\frac{1}{N}$ expansion of $\nu $. This will be used in dealing
with a QH system combining bosons and fermions from the quantum
mechanics point of view. In section 3, we study the supersymmetric
extension of the SP construction by using two approaches:
component fields and world line superfields\footnote{A general
treatment of $\cal N$ = 1 supersymmetry (SUSY) in four-dimensional
can be found in~\cite{gates}. However, a systematic development of
the representation theory of supersymmetric 1 dimensional,
arbitrarily $\cal N$-extended SUSY can be found in~\cite{gates2}.}
By distinguishing two different configurations, we build the
ground states and determine their filling factors. We summarize
our results in the last section.

%%%%%%%%%%%%%%%%%%%%%%%%%%%%%%%%%%%%%%%%%%%%%%%%%
\section{Susskind-Polychronakos matrix model}
%%%%%%%%%%%%%%%%%%%%%%%%%%%%%%%%%%%%%%%%%%%%%%%%%%%

We revisit the SP model by using a method of constrained gauge
systems that we extend to the matrix quantum mechanics. It allows
us to give another way to obtain the filling factor of the system
as well as its $\frac{1}{N}$ expansion. Of course for a large
number of particles $N$ such $O\left( \frac{1}{N}\right) $
corrections are negligible. On the other hand, this approach can
also be applied to the SE of the SP model, which we will develop
in the forthcoming section.

To start recall that the SP action $S\left[ X^{a},\Psi ,A\right]
$, describing a system of $N$ electrons in the presence of a
strong magnetic field $B$, is given by
\begin{eqnarray}
S\left[ X^{a},\Psi ,A_{0}\right] &=&-\frac{B}{2}\int
dt\;\varepsilon _{ab}\; \mathrm{Tr}\left[ \left(
\dot{X}^{a}-i\left[ A,X^{a}\right] \right)
X^{b}+\theta \varepsilon ^{ab}A_{0}  \right]  \notag \\
&&+\int dt\ \left\{\sum_{n,m=1}^{N}i\Psi _{n}^{\ast }\left(
\delta_{nm}
\partial_{t}-iA_{nm}\right) \Psi _{m}-\frac{B\omega }{4}
\; \sum_{a}\mathrm{Tr} \left( X^{a}\right) ^{2}\right\} \label{3}
\end{eqnarray}
where $\theta $ is the non-commutativity parameter, $X^{a}\left(
t\right) $ and $A_{0}=A\left( t\right) $ are one-dimensional (1D)
$N\times N$ hermitian matrices but $\Psi $ is the Polychronakos
field in the fundamental of $U(N)$. Since $X^{a}\left( t\right) $
and $A\left( t\right) $ are hermitian, we may attempt to
diagonalize them by using $U(N)$ invariance of the action and work
directly with the field eigenvalues. This way seems easy and
interesting in quantizing the system, but we will not follow this
path.  Indeed, we will keep $X^{a}\left( t\right) $ and $A\left(
t\right) $ as matrices and impose gauge invariance on the Hilbert
space as usual in constrained systems.

The equations of motion for $X^{a}$ and $\Psi $, respectively, are
\begin{equation}
\begin{array}{l}
\varepsilon _{ab}\nabla _{t}X^{b}+\omega X^{a}=0 \\
\nabla _{t}\Psi _{n}=0
\end{array}
\label{3a}
\end{equation}
where $\nabla _{t}=\partial _{t}-iA$,  There is also an additional
equation of motion for $A$.   Since this latter is one of the key
object in this analysis, we discuss it in detail below.

%%%%%%%%%%%%%%%%%%%%%%%%%%%%%%%%%%%%%%%%%%%
\subsection{Constraint equation}
%%%%%%%%%%%%%%%%%%%%%%%%%%%%%%%%%%%%%%%%%%%

The gauge field $A$ has no kinetic term and can be eliminated via
its equation of motion. This gives the Gauss constraint
\begin{equation}
i\frac{B}{2}\varepsilon _{ab}[X^{a},X^{b}]_{nm}+\Psi _{n}^{\ast
}\Psi _{m}=B\theta \delta _{nm}.  \label{3b}
\end{equation}
Since the hermitian matrices $X^{a}$ are in the $U(N)$ algebra, it
is convenient to rewrite (\ref{3b}) in terms of the
$U(N)=U(1)\oplus SU(N)$ generators $\mathcal{J}_{nm}$
\begin{equation}
\mathcal{J}_{nm}=B\theta \delta _{nm},  \label{3c}
\end{equation}
on the other hand
\begin{equation}
\mathcal{J}_{nm}=i\frac{B}{2}\varepsilon _{ab}\left[
X^{a},X^{b}\right] _{nm}+\Psi _{n}^{\ast }\Psi _{m}.  \label{4}
\end{equation}
Upon setting
\begin{equation}
Z^{\pm }=\frac{1}{\sqrt{B}}\left( X^{1}\pm iX^{2}\right)
\end{equation}
it is easily seen that (\ref{4}) takes the form
\begin{equation}
\mathcal{J}_{nm}=\left[{Z}^+,Z^-\right] _{nm}+\Psi _{n}^{\ast
}\Psi _{m}  \label{4a}
\end{equation}
where $Z=Z^{+}$ and $\overline{Z}=Z^{-}$. Using the $U(N)$
generator basis $\left\{ T^{I},1\leq I\leq N^{2}-1,T^{N^{2}}\equiv
Q\right\} $, with $Q$ is the conserved charge, satisfying the
usual commutation relations
\begin{equation}
\left[ T^{I},T^{J}\right] =if_{K}^{IJ}T^{K}  \label{alcor}
\end{equation}
and expanding $Z$ as
\begin{equation}
Z=\sum_{I}Z_{I}T^{I}  \label{zeq}
\end{equation}
we can show that (\ref{4a}) can be written as
\begin{equation}
\mathcal{J}_{nm}=\sum_{K=1}^{N^{2}-1}\mathcal{J}_{nm}^{K}+\mathcal{J}%
_{nm}^{0}  \label{curent}
\end{equation}
where the components $\mathcal{J}_{nm}^{K}$ are given by
\begin{equation}
\begin{array}{l}
\mathcal{J}_{nm}^{K} =\left( i\overline{Z}_{I}Z_{J}f_{K}^{IJ}+\Psi
^{\ast}T^{K}\Psi \right) T_{nm}^{K}   \\
\mathcal{J}_{nm}^{0} =\left( \Psi ^{\ast }Q\Psi \right) Q_{nm}.
\label{4b}
\end{array}
\end{equation}
Since the first $N^{2}-1$ generators are traceless and
\begin{equation}
\mathrm{Tr}(Q)=N
\end{equation}
 (\ref{4b}) can be simplified as
\begin{equation}
\begin{array}{l}
\mathcal{J}_{nm}^{K} =0  \\
\mathcal{J}_{nm}^{0} =B\theta \delta _{nm}.  \label{4c}
\end{array}
\end{equation}
This can be seen as just the classical conditions for the $SU(N)$
gauge invariance and the fixed total $U(1)$ charge
\begin{equation}
q=\mathrm{Tr}\left( \mathcal{J}_{nm}^{0}\right) =B\theta N.
\label{fixed}
\end{equation}

%%%%%%%%%%%%%%%%%%%%%%%%%%%%%%%%%%%%%%%
\subsection{Quantum analysis}
%%%%%%%%%%%%%%%%%%%%%%%%%%%%%%%%%%%%%%%

The study of constrained gauge systems teaches us that quantum
mechanically the above constrained relations should be written as
\begin{eqnarray}
&&\langle \mathcal{F}|\mathcal{J}_{nm}^{K}|\mathcal{F}\rangle =0
\label{4d}
\\
&&\langle
\mathcal{F}|\mathrm{Tr}\mathcal{J}^{0}|\mathcal{F}\rangle =B\theta
N \label{4e}
\end{eqnarray}
where $|\mathcal{F}\rangle $ are the wave functions of the
underlying
Hilbert space $\mathcal{H}$ of the system. The first equation tells us that $%
|\mathcal{F}\rangle $ should satisfy
\begin{equation}
T_{nm}^{K}|\mathcal{F}\rangle =0  \label{cons}
\end{equation}
and reflects just the $SU(N)$ gauge invariance. Quantization of
the $U(N)$ charges requires that
\begin{equation}
B\theta =kN,\qquad k\in {\mathbb{Z}}_{+}^{\ast }.  \label{4f}
\end{equation}
The second constraint~(\ref{4e}) implies that $|\mathcal{F}\rangle
$ should carry $kN$ $U(1)$ charges.

Furthermore in quantum mechanics with matrix variables, the usual
classical conjugate momentum $\Pi _{\phi }$ and the variables
$\phi $ satisfying the Poisson brackets
\begin{equation}
\left\{ \Pi _{\phi },\phi \right\} _{\mathrm{Poisson}}=1
\label{pbra}
\end{equation}
are roughly speaking interpreted as creation
$\mathbf{O}^{\dagger}$ and annihilation $\mathbf{O}$ operators
obeying generalized Heisenberg commutation relations. For the case
of the square matrix fields $X$ and the Polychronakos vector
variable $\Psi $, we have the relations
\begin{equation}
\begin{array}{l}
\left[ \mathbf{A}_{nm}^{\dagger },\mathbf{A}_{ij}^{-}\right]
=\delta
_{mi}\delta _{nj} \\
\left[ \mathbf{\Psi }_{i}^{\dagger },\mathbf{\Psi }_{j}^{-}\right]
=\delta _{ij}.
\end{array}
\label{5}
\end{equation}
The Hamiltonian $\mathbf{H}$ of the first quantized matrix model
is
\begin{equation}
\mathbf{H}=\omega \left( \mathcal{N}_{\mathrm{A}}+\frac{a_{\mathrm{b}}}{2}%
\right)  \label{6}
\end{equation}
where $\mathcal{N}_{\mathrm{A}}$ is the matrix operator number
\begin{equation}
\mathcal{N}_{\mathrm{A}}=\sum_{n,m}\mathbf{A}_{nm}^{\dagger }\mathbf{A}%
_{nm}^{-}  \label{num}
\end{equation}
and the extra term $\frac{a_{\mathrm{b}}}{2}=\frac{N^{2}}{2}$ is
due to quantum corrections generated by the central term of
(\ref{5}). As we noted previously, in this matrix quantum
mechanical treatment, the result of (\ref{4}) is now replaced by a
constraint on the wave functions $|\mathcal{F}\rangle $ of
$\mathcal{H}$:
\begin{eqnarray}
&&J_{nm}|\mathcal{F}\rangle =0  \label{7} \\
&&J^{0}|\mathcal{F}\rangle =kN|\mathcal{F}\rangle .  \label{8}
\end{eqnarray}
$|\mathcal{F}\rangle $ are invariant under $SU(N)$ and carry a
well defined $J^{0}$ charge. To obtain the solution of these
constraint equations, note that
\begin{equation}
\begin{array}{l}
\left[ J^{0},\mathbf{\Psi }^{\dagger }\right] =\pm \mathbf{\Psi
}^{\dagger }
\\
\left[ J^{0},\mathbf{\Psi }^{\dagger }\mathbf{\Psi }^{-}\right] =0%
\end{array}
\label{com}
\end{equation}
which tell us that the $\mathbf{\Psi }^{\dagger }$ and
$\mathbf{\Psi }^{-}$
dependence in $|\mathcal{F}\rangle $ should be of the form $\left( \mathbf{%
\Psi }_{i}^{\dagger }\right) ^{kN}$ times an arbitrary function of $\mathbf{%
\Psi }^{\dagger }\mathbf{\Psi} ^{-}$, that is $\left( \mathbf{\Psi
}^{\dagger
}\right) ^{kN}\times f\left( \mathbf{\Psi }^{\dagger }\mathbf{\Psi }%
^{-}\right) $. The solution of condition (\ref{7}) leads to the
following normalized ground state
\begin{equation}
|\mathcal{F}_{0}\rangle =\mathcal{N}\left[ \varepsilon ^{\alpha
_{1}\cdots
\alpha _{N}}\;\mathbf{\Psi }_{\alpha _{1}}^{\dagger }\left( \mathbf{\Psi }%
^{\dagger }\cdot \mathbf{A}^{\dagger }\right) _{\alpha _{2}}\cdots
\left(
\mathbf{\Psi }^{\dagger }\cdot \mathbf{A}^{\dagger N}\right) _{\alpha _{N}}%
\right] ^{k}\;|0\rangle.  \label{9}
\end{equation}
One of the remarkable properties of (\ref{9}) is that its
fundamental energy $E_{0}$
\begin{equation}
E_{0}=\frac{\langle \mathcal{F}_{0}|\mathbf{H}|\mathcal{F}_{0}\rangle }{%
\langle \mathcal{F}_{0}|\mathcal{F}_{0}\rangle }  \label{10}
\end{equation}
behaves for large $N$ as
\begin{equation}
\frac{2E_{0}}{\omega }\sim N^{2}\left( k+1\right) .  \label{11}
\end{equation}
Other remarkable features of this solution were considered in \cite%
{hellerman1}.

%%%%%%%%%%%%%%%%%%%%%%%%%%%%%%%%%%%%%%%%%%%%%%%%%
\subsection{Quantum flux}
%%%%%%%%%%%%%%%%%%%%%%%%%%%%%%%%%%%%%%%%%%%%%%%%%%%

This physical quantity is relevant in determining the filling
factor for the SP model, which is defined as the ratio between the
number of particles $N$ and the quantum flux number $N_{\phi }$
\begin{equation}\label{ffd}
\nu =\frac{N}{N_{\phi }}.
\end{equation}
To make contact between this relation and the matrix model
formalism, it is interesting to develop the calculus of $N_{\phi
}$ in terms of the matrix operators of the Hall droplet. This can
be done by considering the classical integral measure
\begin{equation}
d^{2}N_{\phi }\equiv n_{\phi }=Bdx^{1}dx^{2}  \label{11a}
\end{equation}
as the mean value over the $N$ measure integrals
$Bdx_{i}^{1}dx_{i}^{2}$ associated with the coordinates
$x_{i}^{a}$ within the droplet
\begin{equation}
Bdx^{1}dx^{2}=\frac{B}{N}\sum_{i=1}^{N}dx_{i}^{1}dx_{i}^{2}=\frac{1}{N}%
\sum_{i=1}^{N}dx_{i}d\pi _{i}  \label{11b}
\end{equation}
where $x_{i}$ and $\pi _{i}$ stand for $x_{i}^{1}$ and
$Bx_{i}^{2}$\ respectively. Now using the fact that $x_{i}^{a}$
variables are nothing but
the eigenvalues of the hermitian matrices $X$ and $\Pi $, one can rewrite (%
\ref{11a}) as
\begin{equation}
n_{\phi }=\frac{1}{2N}\mathrm{Tr}\left( dXd\Pi +d\Pi dX\right)
\label{12}
\end{equation}
where $X$ ($X=X^{1}$) and its conjugate momentum $\Pi $ ($\Pi
=BX^{2}$) are representants of the $U(N)$ gauge orbit
\begin{equation}
X\equiv UXU^{\dagger },\qquad \Pi \equiv U\Pi U^{\dagger },\qquad
U\in U\left( N\right) .  \label{12a}
\end{equation}
With these matrix variables, the classical relation for the flux
integral
\begin{equation}
N_{\phi }=\int_{\mathrm{droplet}}Bdx^{1}dx^{2}
\end{equation}
can be written as
\begin{equation}
N_{\phi }=\int_{\mathrm{droplet}}\frac{1}{N}\mathrm{Tr}\left(
dXd\Pi \right) =\frac{1}{2N}\mathrm{Tr}_{\mathcal{H}_{\rm
droplet}}\left( X\Pi +\Pi X\right) \label{13}
\end{equation}
where $\mathcal{H}_{\mathrm{droplet}}$ stands for the Hilbert
space of the droplet. This is a remarkable relation first because
it expresses the number $N_{\phi }$ in terms of the coordinates of
the phase space of the SP model and second because of its
interpretation at the quantum level. By using an analysis similar
to that we have adopted for (\ref{4}), we can also obtain
\begin{equation}
N_{\phi }=\frac{1}{N}\sum_{|\mathcal{F}\rangle \in
\mathcal{H}}\langle \mathcal{F}|\mathcal{N}_{\phi
}|\mathcal{F}\rangle  \label{131}
\end{equation}
where the flux number operator $\mathcal{N}_{\phi }$ that follows
from (\ref{13}) is given by
\begin{equation}
\mathcal{N}_{\phi }=\frac{1}{2}\sum_{n,m}\left( \mathbf{A}_{nm}^{\dagger }%
\mathbf{A}_{nm}^{-}+\mathbf{A}_{nm}^{-}\mathbf{A}_{nm}^{\dagger
}\right) . \label{14}
\end{equation}
Now using (\ref{5}) and (\ref{num}), we can show that
\begin{equation}
\mathcal{N}_{\phi }=\mathcal{N}_{\mathrm{A}}+\frac{N^{2}}{2}
\end{equation}
which implies in turn that for the SP model there is a close relation between $%
\mathcal{N}_{\phi }$ and $\mathbf{H}$, namely
\begin{equation}
\mathbf{H}=\omega \mathcal{N}_{\phi }  \label{141}
\end{equation}
or equivalently by using (\ref{131})
\begin{equation}
E=\frac{\omega }{2}N_{\phi }N.  \label{142}
\end{equation}
This tells us that the energy spectrum of the present system can
be measured in terms of the quantum flux number and the number of
electrons. Note also that for a fixed number of particles $N$,
energy variation is induced by the variation of the flux.

%%%%%%%%%%%%%%%%%%%%%%%%%%%%%%%%%%%%%%%%%%%%%%%%%%%%%%%%%%%%%%%%%%%%%%%%%%%%
\subsection{$\mathcal{O}\left( 1/N\right) $ expansion of the filling factor}
%%%%%%%%%%%%%%%%%%%%%%%%%%%%%%%%%%%%%%%%%%%%%%%%%%%%%%%%%%%%%%%%%%%%%%%%%%%%

Equation (\ref{142}) is an interesting result since it has three
different consequences. First, it relates the filling factor
$\nu$~(\ref{ffd}) to the energy of the states. Indeed,
substituting
\begin{equation}
N_{\phi }=\frac{2E}{\omega N}
\end{equation}
we end up with
\begin{equation}
\nu =\frac{\omega N^{2}}{2E}.  \label{fo}
\end{equation}
For the lowest Landau level of energy $E_{0}$, the quantum flux is
\begin{equation}
 N_{\phi }\sim\frac{2E_{0}}{\omega N}
\end{equation}
 and $\nu$ reduces to the familiar formula
 \begin{equation}
\nu _{0}=\frac{\omega N^{2}}{2E_{0}}.
\end{equation}

At this stage, we believe a remark is in order. As we know, it is
not so easy to solve interacting QH systems. Nevertheless, one can
always use some other technique to acquire more information about
such systems. For instance, in the perturbation theory the energy
$E$ can be expanded as
\begin{equation}
E=E_{0}\left( 1+\frac{E_{1}}{E_{0}}+\frac{E_{2}}{E_{0}}+\cdots
\right)
\end{equation}
where $E_{i}$ are the perturbated energies and
$\frac{E_{i}}{E_{0}}$ are smaller than one.  Now (\ref{fo})
becomes
\begin{equation}
\nu =\frac{\omega N^{2}}{2E_{0}}\left( 1+\frac{E_{1}}{E_{0}}+\frac{E_{2}}{%
E_{0}}+\cdots \right) ^{-1}
\end{equation}
which suggests that one may expand it in a perturbative series
\begin{equation}
\nu =\sum_{n\geq 0}\nu _{n}
\end{equation}
with the leading term given by $\nu _{0}$ of the ground state. For
a strongly correlated system with a large gap, particles are
mainly confined in the lowest Landau level. In this case, by using
(\ref{141}), we can show that (\ref{131}) reduces to
\begin{equation}
N_{\phi }\sim \frac{1}{N}\langle \mathcal{F}_{0}|\mathcal{N}_{\mathrm{A}}|%
\mathcal{F}_{0}\rangle +\frac{N}{2}.  \label{15}
\end{equation}
With the help of
\begin{equation}
\mathcal{N}_{\mathrm{A}}|\mathcal{F}_{0}\rangle
=\frac{kN}{2}\left( N+1\right) |\mathcal{F}_{0}\rangle
\label{acting}
\end{equation}
which can be seen from (\ref{9}), we obtain
\begin{equation}
N_{\phi }\sim k+N\left( k+1\right).  \label{16}
\end{equation}
This leads to the well-known filling factor
\begin{equation}
\nu _{0}=\frac{1}{k+1}  \label{fff}
\end{equation}
which is in agreement with \cite{polychronakos, hellerman1}.

A second comment concerns the general expression for $E$ and
$N_{\phi }$. From the way of building up the SP wave functions
$|\mathcal{F}\rangle $, we note that the expectation value
\begin{equation}
\langle \mathcal{F}|\mathcal{N}_{\mathrm{A}}|\mathcal{F}\rangle
\end{equation}
can be usually written as a power series in terms of the integer
$N$. This implies that the total energy $E$ of the SP approach can
be expanded as
\begin{equation}
E=\frac{\omega }{2}\sum_{s\geq 1}r_{s}N^{s}  \label{161}
\end{equation}
where $r_{s}$ are rational numbers whose two leading terms are
\begin{equation}
r_{1}=k, \qquad r_{2}=k+1.
\end{equation}
Now by comparing (\ref{142}) and (\ref{161}), we obtain
\begin{equation}
N_{\phi }=\sum_{s\geq 1}r_{s}N^{s-1}=\frac{1}{N}\left(
r_{1}N+r_{2}N^{2}+r_{3}N^{3}+\cdots \right) .  \label{162}
\end{equation}
With this relation, one can compute the finite part of the ratio $\frac{%
N_{\phi }}{N}$ to obtain the general form of the filling factor in
such way that
\begin{equation}
\nu =\frac{1}{k+1}-\frac{k}{N(k+1)}-\left( \frac{k}{N(k+1)}\right) ^{2}+%
\mathcal{O}\left( \frac{1}{N^{3}}\right)  \label{163}
\end{equation}
which is a $1/N$ expansion series. Note that it converges to $ \nu
_{0}$ (\ref{fff}) in the limit $N\rightarrow \infty $. The extra
terms can be interpreted as $1/N$ quantum fluctuations.

A third aspect upon which we wish to comment concerns another
property of (\ref{fo}). In general, this relation may be applied
to the SE of the SP model, which we will study later. In the
absence of mutual interactions between bosons and fermions, the
total energy $E$ of the supersymmetric QH system is the sum of two
parts
\begin{equation}
E_{\mathrm{bf}}=E_{\mathrm{b}}+E_{\mathrm{f}}
\end{equation}
where $\mathrm{b}$ and $\mathrm{f}$ refer, respectively, to bosons
and fermions. Then by adopting the above analysis one expects that
the corresponding filling factor is given by
\begin{equation}
\nu _{\mathrm{bf}}=\frac{\omega N^{2}}{2\left( E_{\mathrm{b}}+E_{\mathrm{f}%
}\right) }.  \label{fa}
\end{equation}

%%%%%%%%%%%%%%%%%%%%%%%%%%%%%%%%%%%%%%%%%%%%%%%%
\section{Supersymmetric matrix model}
%%%%%%%%%%%%%%%%%%%%%%%%%%%%%%%%%%%%%%%%%%%%%%%%

Generally speaking a particle confined to a plane in the presence
of a strong perpendicular magnetic field $B$ behaves as a 1D field
$x(t)$ $(\equiv x^{1})$ with conjugate momentum\footnote{ For a
system of $N$ particles, the conjugate momenta $x_{i}^{2}\left(
t\right) $ are thought of as the eigenvalues of a $N\times N$
hermitian matrix field $X^{2}\left( t\right) $. The $X^{1}$ and
$X^{2}$ matrices satisfy the Heisenberg algebra with central
extension proportional to the inverse of the external
magnetic field.} $y\left( t\right) $ $%
(\equiv x^{2})$. The theories of such configurations are quite
well understood and the matrix model of the previous section is
one of them.  For a system combining bosons and fermions, one may
have an interesting situation that deserves to take care of. A
generic configuration of such a system can be imagined as
consisting of two phases: U  involving $N_{1}$ bosons and D
involving $N_{2}$ fermions. To fix the ideas, we take these
numbers as
\begin{equation}
N_{1}=k_{1}N,\qquad N_{2}=k_{2}N
\end{equation}
where $k_{1}$ and $k_{2}$ are positive integers. Note that for
$k_{2}=0$, we have no D phase and the QH system is the one
described in previous section. For non-zero $k_{2}$, the situation
is important in the sense that is quite realistic. In particular,
$k_{2}=1$ is interesting because it has much to do with the SE of
the SP model. Therefore, we claim that our supersymmetric
generalization of the SP theory modelled the U and D
configurations with a positive odd integer $k_{1}$ and $k_{2}=1$.

%%%%%%%%%%%%%%%%%%%%%%%%%%%%%%%%%%%%%%%%%%%%%%%%%%%%%%%%%%%%%%
\subsection{Generalization to supersymmetric matrix theory}
%%%%%%%%%%%%%%%%%%%%%%%%%%%%%%%%%%%%%%%%%%%%%%%%%%%%%%%%%%%%%%

We start by recalling that in the SMM we have essentially the same
tools and machinery as those we
generally have in the SP model. The main difference is that %usual
the usual QH system is now replaced by another system describing
multiplets involving bosonic and fermionic states. The main
novelties in the SMM are of two types:

{\bf(i)} This model describes two kinds of the fractional QH
subsystems: bosons and fermions (superpartner) forming,
respectively, U and D phases. Classically, they are described by
two pairs of 1D fields namely the super-position coordinates
\begin{equation}
x_{i}^{1}\left( t\right) ,\qquad \eta _{i}^{1}\left( t\right)
\label{m1}
\end{equation}
interchanged under supersymmetric transformations (ST) and their
respective conjugate momenta
\begin{equation}
x_{i}^{2}\left( t\right) ,\qquad \eta _{i}^{2}\left( t\right) .
\label{m2}
\end{equation}
They are related by $1D$ supersymmetry (SUSY) as
\begin{equation}
\delta _{\epsilon }\eta ^{a}=\epsilon x^{a},\qquad \delta
_{\epsilon }x^{a}=-i\epsilon \partial _{t}\eta ^{a}  \label{m3}
\end{equation}
where $\epsilon $ is a parameter of the SUSY group with $\epsilon
^{2}=0$ and $a=1,2$. Note that the ST we are using here is the
world line SUSY. It should be distinguished from the target space
SUSY which might also be considered for building more general
super QH systems with rich invariance. We will comment later on
this issue, but for the moment we will fix our attention mainly on
our world line SMM.

{\bf(ii)} In the SMM formulation, the multiplets (\ref{m1}) and
(\ref{m2}) are replaced by a pair of 1D $N\times N$
 hermitian matrix fields
 $X^{a}=X^{a}\left( t\right) $ and $\eta ^{a}=\eta
^{a}\left( t\right) $
\begin{equation}
X^{a}=\left(
\begin{array}{cccc}
X_{11}^{a} & . & . & X_{1N}^{a} \\
. & . & . & . \\
. & . & . & . \\
X_{N1}^{a} & . & . & X_{NN}^{a}%
\end{array}
\right) ,\qquad {\ \eta }^{a}=\left(
\begin{array}{cccc}
{\eta }_{11}^{a} & . & . & {\ \eta }_{1N}^{a} \\
. & . & . & . \\
. & . & . & . \\
{\ \eta }_{N1}^{a} & . & . & {\ \eta }_{NN}^{a}%
\end{array}
\right) .  \label{m4}
\end{equation}
Since these matrices are hermitian, they can be diagonalized by
making use of the $U(N)$  %\newline  $~\,\,\,~~$ $~\,\,~\,~$
 transformations in such way that
\begin{equation}
\begin{array}{l}
V_{ik}X_{kl}^{a}V_{kj}^{-1}=X_{i}^{a}\delta _{ij} \\
U_{ik}\eta _{kl}^{a}U_{kj}^{-1}=\eta _{i}^{a}\delta _{ij}.%
\end{array}
\label{eq}
\end{equation}
As in the bosonic case, we will not fix such a symmetry at the
level of the dynamical matrix field variables. We will use it as a
basic symmetry in the action and only impose it at the level of
the Hilbert space of wave functions describing the fractional QH
system. Since we are going to consider large but still finite
$N\times N$ matrices, we need to introduce the Polychronakos
vector field $\Psi =\Psi \left( t\right) $ as well as its
superpartner $\upsilon =\upsilon \left( t\right) $
\begin{equation}
\Psi =\left(
\begin{array}{cccc}
\Psi _{1}, & . & . & ,\Psi _{N}%
\end{array}
\right) ,\qquad \upsilon =\left(
\begin{array}{cccc}
\upsilon _{1}, & . & . & ,\upsilon _{N}%
\end{array}
\right) .  \label{m5}
\end{equation}
%$~\,~\,~$ $~\,\,~\,~$
To write the action of the world line SMM, we may use two
different but equivalent ways: component field and world line
superfield approaches. The latter involves auxiliary fields and is
more general. For completeness, we shall consider in this paper
both of them.

%%%%%%%%%%%%%%%%%%%%%%%%%%%%%%%%%%%%%%%%%
\subsection{Component field method}
%%%%%%%%%%%%%%%%%%%%%%%%%%%%%%%%%%%%%%%%%

One way to build the world line SMM is to use a direct method
based on component field techniques. This allows us to derive
immediately the world line SE of the constraint equation
(\ref{3b}) and write down the corresponding supersymmetric
invariant action. This can be done by considering the following ST
for matrix variables\footnote{ In presence of gauge fields,
$\partial _{t}$ should be replaced by $\nabla _{t}=\partial
_{t}-iA_{t}$.}
\begin{equation}
\begin{array}{l}
\delta _{\epsilon } {\eta }^{a}=\epsilon X^{a},\qquad \delta
_{\epsilon
}X^{a}=-i\epsilon \partial _{t}{\eta }^{a} \\
\delta _{\epsilon }\upsilon =\epsilon \Psi ,\qquad \delta
_{\epsilon }\Psi =-i\epsilon \partial _{t}\upsilon.
\end{array}
\label{16a}
\end{equation}
They extend (\ref{m3}) and satisfy the following $1D$
$\mathcal{N}=1$ supersymmetric algebra
\begin{equation}
\left[ \delta _{\epsilon _{1}},\delta _{\epsilon _{2}}\right] \phi
=2i\epsilon _{1}\epsilon _{2}\partial _{t}\phi  \label{susyal}
\end{equation}
where $\phi $ stands for $X^{a},\eta ^{a},\Psi $ and $\upsilon $.
Before going on, let us list the contents of supersymmetric
multiplets, which will be used. These are of two kinds:

\textbf{(i)} $N^{2}$ Susskind supermultiplets $\left(
0^{N^{2}},\left( 1/2\right)^{N^{2}}\right) $ described by $\left(
X_{ij},\eta _{ij}\right) $ matrix field %\newline  $~\,\,\,~\,~$ $~\,~~$
transforming in the $U\left(N\right) $ adjoint representation.

\textbf{(ii)} $N$ supermultiplets $\left( \left( -1/2\right)
^{N},0^{N}\right) $, $\left( -1/2\right) ^{N}$ for the
Polychronakos superpartner, described by $\left( \Psi
_{i},\upsilon _{i}\right) $ transforming as the vector of $U\left(
N\right) $. Since $X^{a}$ and $\eta ^{a}$ are hermitian, they can
be diagonalized as $\left( X_{i}\delta _{ij},\eta _{i}\delta
_{ij}\right) $ and so one has a $N$-dimensional target space
geometry. Otherwise the dimension would be $N^{2}$ due to the
equivalence (\ref{eq}).

%%%%%%%%%%%%%%%%%%%%%%%%%%%%%%%%%%%%%%%%%%%%%%%%%%%%%%%%%%%%%%%%%%%%%%
\subsubsection{Supersymmetric constraint equations}
%%%%%%%%%%%%%%%%%%%%%%%%%%%%%%%%%%%%%%%%%%%%%%%%%%%%%%%%%%%%%%%%%%%%%%

We propose to study the SE of the constraint equations by
considering classical and quantum analysis.

\textbf{(1) Classical constraints}: From the above transformations
together with the gauge condition $\delta _{\epsilon }A=0$ (which
in a complete supersymmetric formulation should involve a
supersymmetric partner: gauginos) one can build the supersymmetric
generalization of the SP theory just by demanding closure under
(\ref{16a}). According to (\ref{3a}), we find the $X$ and $\Psi $
superpartner equations of motion
\begin{equation}
\begin{array}{l}
\varepsilon _{ab}\nabla _{t}^{2}\eta ^{b}+\omega \nabla _{t}\eta ^{a}=0 \\
\nabla _{t}^{2}\upsilon _{n}=0.
\end{array}
\label{16A}
\end{equation}
Assuming that these equations result from the variation of an
action $S_{\rm susy}=S\left[ \phi \right] $, i.e.
\begin{equation}
\delta S/\delta \phi =0
\end{equation}
one can derive the SE of
the SP action (\ref{3}). Doing this integration calculus, we end
up with
\begin{eqnarray}
S_{\rm susy} &=&-\frac{B}{2}\int dt\varepsilon _{ab}\
\mathrm{Tr}\left( \nabla _{t}X^{a}X^{b}+\frac{i}{2}\nabla _{t}\eta
^{a}\nabla _{t}\eta ^{b}+\theta
\varepsilon ^{ab}A_{0}\right)  \notag \\
&&+\int dt\ \left\{i\Psi ^{\ast }\nabla _{t}\Psi +\nabla
_{t}\upsilon^{\ast}\nabla _{t}\upsilon -\frac{B\omega
}{4}\;\sum_{a}\mathrm{Tr}\left( X^{a}X^{a}-i\eta ^{a}\nabla
_{t}\eta ^{a}\right) \right\} \label{16B}.
\end{eqnarray}
We can show that $S_{\rm susy}$ has the $U\left( N\right) $ gauge
invariance of the bosonic SP action. Actually we have two
conserved currents $\mathcal{J}_{mn}$ and $\mathcal{F}_{mn}$
valued in $U\left( N\right) $. They, respectively, generate
bosonic and fermionic symmetries and combine to form a unique
supercurrent in superspace formalism. $\mathcal{J}_{mn}$ has two
contributions: bosonic $\mathcal{J}_{({\mathrm{b}})mn}$ and
fermionic $\mathcal{J}_{({\mathrm{f}})mn}$ and can be expanded in
the $u\left( N\right) $ basis $\left\{ T^{K},1\leq K\leq
N^{2}\right\} $ as
\begin{equation}
\mathcal{J}_{mn}=\mathcal{J}_{({\mathrm{b}})mn}+
\mathcal{J}_{({\mathrm{f}})mn}=\sum_{K=1}^{N^{2}}T^{K}\mathcal{J}_{nm}^{K}.
\label{jcur}
\end{equation}
In a similar way we can do so for $\mathcal{J}_{(\mathrm{{b})mn}}$
and $\mathcal{J}_{(\mathrm{{f})mn}}$.
$\mathcal{J}_{(\mathrm{{f})mn}}$ is completely new as it is
generated by the world line fermions and has no analogue in the
bosonic SP matrix model. The 1D fermionic current
$\mathcal{F}_{mn}$ can also be expanded as
\begin{equation}
\mathcal{F}_{mn}=\sum_{K=1}^{N^{2}}T^{K}\mathcal{F}_{nm}^{K}.
\label{fcur}
\end{equation}
Note that $\mathcal{J}_{mn}$ and $\mathcal{F}_{mn}$ are not
completely independent since they are related under the ST
\begin{eqnarray}
\delta _{\epsilon }\mathcal{F}_{nm} &=&\epsilon \mathcal{J}_{mn}  \notag \\
\delta _{\epsilon }\mathcal{J}_{nm} &=&-i\epsilon \partial _{t}\mathcal{F}%
_{mn}.  \label{16b}
\end{eqnarray}
and (\ref{susyal}). Their explicit component field expressions,
defining the classical constraint equations, are given by
\begin{eqnarray}
&&\mathcal{F}_{mn}=i\frac{B}{2}\varepsilon _{ab}\left[ {\eta
}^{a},X^{b} \right]_{nm}+\frac{1}{2}\left( \upsilon _{n}^{\ast
}\Psi _{m}+\Psi
_{m}^{\ast }\upsilon _{n}\right)  \label{16c1} \\
&&\mathcal{J}_{({\rm b}) mn}=i\frac{B}{2}\varepsilon _{ab}\left[
X^{a},X^{b}\right] _{nm}+\Psi _{n}^{\ast }\Psi _{m}  \label{16c} \\
&&\mathcal{J}_{({\rm f}) mn}=-\frac{B}{2}\varepsilon _{ab}\left\{
{ \eta }^{a},\partial _{t}{ \eta }^{b}\right\}
_{nm}+\frac{i}{2}\left( \upsilon _{n}^{\ast }\partial _{t}\upsilon
_{m}-\upsilon _{n}\partial _{t}\upsilon _{m}^{\ast }\right) .
\label{16c3}
\end{eqnarray}
These quantities can be rewritten in terms of the $U\left(
1\right) \oplus SU\left( N\right) $ generators in a fashion
similar to (\ref{4b}-\ref{4c}). The new thing, which appears in
comparison to the SP theory, is that these generators have two
sectors in one-to-one correspondence with the particles and their
super-partners. For instance, the total $U(1)$ charge $\mathcal{J}
^{0}=\mathrm{Tr}\left( \mathcal{J}_{mn}\right) $ has now two
contributions $\mathcal{J}_{({\rm b}) mn}\equiv\mathcal{J}_{\Psi
}^{0}$ and $\mathcal{J}_{({\rm f}) mn}\equiv \mathcal{J}_{\upsilon
}^{0}$. The old
\begin{equation}
\mathcal{J}_{\Psi }^{0}=N\left( \Psi ^{\ast }Q\Psi \right)
\label{jcontr}
\end{equation}
is realized in terms of the Polychronakos bosonic field $\Psi $ as
in (\ref {16c}) and the extra contribution
\begin{equation}
\mathcal{J}_{\upsilon }^{0}=\frac{i}{2}N\left( \upsilon ^{\ast
}Q\partial _{t}\upsilon -\upsilon Q\partial _{t}\upsilon ^{\ast
}\right) \label{jzcont}
\end{equation}
comes from the world line $\upsilon $ fermions (\ref{16c3}).
Therefore, in addition to the $SU(N)$ invariance, the constraint
relation regarding the $ U\left( 1\right) $ charge reads
\begin{equation}
\mathcal{J}^{0}=\mathcal{J}_{\Psi }^{0}+\mathcal{J}_{\upsilon
}^{0}=Nk \label{16d}
\end{equation}
where its two parts are
\begin{equation}
\mathcal{J}_{\Psi }^{0}=Nk_{1},\qquad \mathcal{J}_{\upsilon
}^{0}=Nk_{2} \label{16e}
\end{equation}
implying that
\begin{equation}
k_{1}+k_{2}=k.
\end{equation}
With this splitting, one can  compare at any step with what we
know about the world line bosonic case and learn exactly what is
the contribution of the world line fermions.

\textbf{(2) Quantum constraints}: In this case the classical
constraint equations (\ref{16c1}-\ref{16e}) should be replaced by
those for the Hilbert space of the SMM
\begin{eqnarray}
&&\langle \mathcal{F}|\left[ \mathcal{J}_{\Psi}^{K}+\mathcal{J}
_{\upsilon}^{K}\right] T_{mn}^{K}|\mathcal{F}\rangle =0
\label{16f}
\\
&&\langle \mathcal{F}|\mathcal{J}^{0}|\mathcal{F}\rangle = \langle
\mathcal{F}|\mathcal{J}_{\Psi }^{0}+\mathcal{J}_{\upsilon
}^{0}|\mathcal{F}\rangle =Nk_{1}+Nk_{2}  \label{16g} \\
&&\langle
\mathcal{F}|\mathcal{F}^{K}T_{mn}^{K}|\mathcal{F}\rangle=0
\label{16fg}
\end{eqnarray}
where $K=1,\cdots , N^{2}-1$ and the state $|\mathcal{F} \rangle $
is a generic wave function of the underlying Hilbert space $
\mathcal{H}$ of the Hamiltonian $\mathbf{H}_{\rm susy}$ of the
world line SMM. As in the SP model, the corresponding
$\mathbf{H}_{\rm susy}$ operator, which may also be written in
terms of the total quantum flux operator as $ \omega
\mathcal{N}_{\phi }$ (\ref{141}), plays a crucial role in the SMM.
It has two operator contributions, one coming from the world line
bosonic operators and the other from their superpartners. It reads
\begin{equation}
\mathbf{H}=\omega \left(
\mathcal{N}_{\mathrm{A}}+\mathcal{N}_{\mathrm{C}}+
\frac{a_{\mathrm{b}}+a_{\mathrm{f}}}{2}\right)  \label{16i}
\end{equation}
where
\begin{equation}
\mathcal{N}_{\mathrm{A}}=\sum_{n,m}\mathbf{A}_{nm}^{\dagger
}\mathbf{A} _{nm}^{-},\qquad
\mathcal{N}_{\mathrm{C}}=\sum_{n,m}\mathbf{C}_{nm}^{\dagger
}\mathbf{C}_{nm}^{-} \label{opnu}
\end{equation}
are, respectively, the number matrix operators counting the world
line bosonic and fermionic excitations. The number
$a_{\mathrm{b}}+a_{\mathrm{f}}$ is the total quantum correction
generated by the ordering of the (world line bosonic and
fermionic) creation and annihilation matrix operators $\mathbf{A}
_{nm}^{\pm }$ and $\mathbf{C}_{nm}^{\pm }$. In general it is equal
to
\begin{equation}
a_{\mathrm{b}}+a_{\mathrm{f}}=N_{\mathrm{e}}^{2}-N_{\widetilde{\mathrm{e}}%
}^{2}
\end{equation}
and cancels exactly due to the world line SUSY, which requires an
equal number of bosonic and fermionic degrees of freedom
\begin{equation}
N_{\mathrm{e}}=N_{\widetilde{\mathrm{e}}}.
\end{equation}
There is nothing strange with this kind of cancellations. It is
well-known in supersymmetric quantum field theories. But the
novelty here is that in the SMM, one expects that there will be no
Polychronakos quantum effect for the total filling factor, the
bosonic and fermionic effects cancel each other. We will turn to
this special feature later, for the moment let us build up the
ground state of our model.

%%%%%%%%%%%%%%%%%%%%%%%%%%%%%%%%%%%%%%%%%%%%%%%%%%%%%%%%%
\subsubsection{Solving quantum constraints}
%%%%%%%%%%%%%%%%%%%%%%%%%%%%%%%%%%%%%%%%%%%%%%%%%%%%%%%%%

We begin by giving some useful tools in order to derive the
lowest-energy wave function of our model. As our system involves
two kinds of operators, the corresponding Hilbert space is a
product of the bosonic and fermionic sectors such as
$\mathbb{F}_{\mathrm{b}}\times \mathbb{F}_{\mathrm{f}}$. They
involve, respectively, the bosonic $\left(\mathbf{A}_{nm}^{\dagger
},\mathbf{A}_{mn}^{-}\right) $ and fermionic $\left(
\mathbf{C}_{nm}^{\dagger },\mathbf{C}_{mn}^{-}\right) $ operators
acting on $|0\rangle $ and $|S\rangle $ vacua. Since the analysis
of the first sector has been discussed in section 2, let us
discuss below that dealing with quantum world line
fermions\footnote{ A priori, the quantization of world line
fermions may have  Neveu--Schwartz (NS) and Ramond (R) like
sectors in analogy with two-dimensional conformal field theory.
Here we focus on the R sector, but a complete analysis would
involve all sectors. It is in this way that one may consider
target space supersymmetric fractional QH models.}.

\textbf{(1)} \textbf{Fermionic oscillators:} Note that on a
generic basis, the harmonic oscillator matrix operators $\left(
\mathbf{C} _{kl}^{\dagger },\mathbf{C}_{nm}^{-}\right) $ form a
system of $N^{2}$ fermionic (Ramond like sector) creation and
annihilation operators. They are the superpartners of the world
line bosonic operators $\left( \mathbf{A}_{nm}^{\dagger
},\mathbf{A}_{mn}^{-}\right) $ of the SP model and satisfy the
Clifford algebra
\begin{eqnarray}
\left\{ \mathbf{C}_{kl}^{\dagger },\mathbf{C}_{nm}^{-}\right\}
&=&-i\delta
_{kn}\delta _{lm} \notag \\
\left\{ \mathbf{C}_{kl}^{\dagger },\mathbf{C}_{nm}^{\dagger
}\right\} &=&\left\{
\mathbf{C}_{kl}^{-},\mathbf{C}_{nm}^{-}\right\} =0.  \label{16j}
\end{eqnarray}
where the different indices are running from $1$ to $N$. It is
known that this algebra has a spinor representation of dimension
$2^{\left[ \frac{N^{2}}{2}\right] }$. Therefore the ground state
$|S\rangle $ of the fermionic field Fock space $\mathbb{F}_{
\mathrm{f}}$ is a $2^{\frac{N^{2}}{2}}$ dimensional spinor in
contrary to the bosonic Fock space $\mathbb{F}_{\mathrm{b}}$ where
the vacuum $|0\rangle $ is 1D. Note that the degeneracy in
$|S\rangle $ is not relevant from the world line SUSY view. It
might be interesting in building a target space supersymmetric
fractional QH matrix theory, but this generalization goes beyond
the scope of the present study. The representations of (\ref{5})
and (\ref{16j}) imply
\begin{eqnarray}
&&\mathbf{C}_{nm}^{-}|S\rangle =0,\qquad {\mathbf H}_{\rm susy}
|S\rangle =-\frac{\omega }{2}N^{2}|S\rangle  \label{e1} \\
&&\mathbf{A}_{nm}^{-}|0\rangle =0,\qquad {\mathbf H}_{\rm susy}
|0\rangle =\frac{\omega }{2}N^{2}|0\rangle  \label{e2} \\
&&\mathbf{A}_{mn}^{\dagger }|0\rangle =|1\rangle ,\qquad \mathbf{C}%
_{mn}^{\dagger }|S\rangle =|S^1\rangle  \label{e3}
\end{eqnarray}
where $|S^1\rangle $ and $|1\rangle $ are the first excited states
of the fermionic and bosonic oscillators. Their respective
energies are $\frac{\omega }{2}\left(1-N^{2}\right) $ and
$\frac{\omega }{2}\left( 1+N^{2}\right) $. Generic states of the
tensor product Hilbert space of the world line SMM are generated
by the product of the following states
\begin{equation}
\prod_{i\geq 1}^{R_{\mathrm{f}}}\mathbf{C}_{k_{i}l_{i}}^{\dagger
}|S\rangle ,\qquad \prod_{i\geq
1}^{R_{\mathrm{b}}}\mathbf{A}_{k_{i}l_{i}}^{\dagger }|0\rangle
\label{e4}
\end{equation}
where $R_{\mathrm{b}}$ and $R_{\mathrm{f}}$ are positive integers.
Of course the first excited states (\ref{e3}) are not the general
ones and (\ref{e4}) cannot be solutions for the quantum
supersymmetric constraints (\ref{16f}-\ref{16g}). The point is
that these states are not $SU\left( N\right) $ gauge invariant and
do not have the appropriate $U\left( 1\right) $ group charge
$k_{1}N+k_{2}N$. To find the appropriate solutions, we proceed as
follows.

\underline{Extra restriction and D phase}: Note that $R_{%
\mathrm{b}}$ appearing in the generating monomials (\ref{e4}) can
take any positive integer value, but $R_{\mathrm{f}}$ should not
exceed $N^{2}$ because of the world line fermion statistics.
Indeed from (\ref{16j}), one can see that the individual
$\mathbf{C}_{kl}^{\dagger }$ operators are nilpotent
\begin{equation}
\left( \mathbf{C}_{kl}^{\dagger }\right) ^{2}=0
\end{equation}
and therefore the non-zero wave functions should have no more than $%
\prod_{i,j=1}^{N}\mathbf{C}_{ij}^{\dagger }$ fermionic
excitations. Otherwise they are nilpotent because
\begin{equation}
\mathbf{C}_{kl}^{\dagger }\left(
\prod_{i,j=1}^{N}\mathbf{C}_{ij}^{\dagger }\right) =0.  \label{c0}
\end{equation}
This means that the wave functions for our model, which we are
looking for, should have no more than $N^{2}$ world line fermions,
but can have any number of world line bosons. The following
proposition summarizes this particular property and makes a link
with the U and D phases.

\textbf{(2) Proposition:} The $U\left( N\right) $ gauge invariant
ground state configuration of the world line SMM describing a set
of $N$ supermultiplets in an external magnetic field $B$ has two
phases:

{\bf(a)} The world line boson phase U corresponds in standard
fractional QH theory to that system described by the SP theory.
Therefore the flux number is given by
\begin{equation}
N_{\phi }^{\mathrm{b}}=\left( k_{1}+1\right) N
\end{equation}
leading to
\begin{equation}
\nu _{\mathrm{b}}=\frac{1}{k_{1}+1}  \label{bff1}
\end{equation}
where $k_{1}$ is a (odd) positive integer. This result agrees with
the Polychronakos effect.

{\bf(b)} The world line fermion phase D  has the flux number
\begin{equation}
N_{\phi }^{\mathrm{f}}=\left( 1-k_{2}\right) N
\end{equation}
where $k_{2}$ restricted to the values $0$ and $1$ in agreement
with the Pauli exclusion principle. The corresponding filling
factor reads
\begin{equation}
\nu _{\mathrm{f}}=\frac{1}{1-k_{2}}.  \label{fff1}
\end{equation}
It shows that the quantum corrections generate an
anti-Polychronakos effect. This is an interesting feature since it
will play a crucial role in characterizing the QH system.

{\bf(c)} From the above analysis, we conclude that the total
filling factor $\nu _{\mathrm{tot}}$ of the supersymmetric
fractional QH system is
\begin{equation}
\nu _{\mathrm{tot}}=\frac{\nu _{\mathrm{b}}\nu _{\mathrm{f}}}{\nu _{\mathrm{b%
}}-\nu _{\mathrm{f}}}.
\end{equation}
It is clear that the Polychronakos and anti-Polychronakos effects
are equal. Since they are with opposite sign, they cancel each
other.

To prove this proposition, we need the ground state configuration.
This can be obtained by solving the quantum supersymmetric
constraints, which requires $U(N)$ gauge invariant wave functions.
Since we have $U(N)=U(1)\times SU(N)$, we will first solve the
constraint equations for the $U(1)$ subsymmetry and second those
for $SU(N)$.

\underline{$U(1)$ symmetry and extra oscillators}: Recall that in
the bosonic case, the $U\left( 1\right) $ gauge invariance is
solved with the help of the Polychronakos vector operators $\left(
\mathbf{\Psi}_{i}^{\dagger },\mathbf{\Psi }_{i}^{-}\right) $
 as shown in (\ref{9}). The same thing happens for
the supersymmetric case. Instead of the Polychronakos operators,
the job will be done by the supersymmetric pairs $\left(
\mathbf{\Psi } _{i}^{\dagger },\mathbf{\upsilon }_{i}^{\dagger
}\right) $ and $\left( \mathbf{\Psi
}_{i}^{-},\mathbf{\upsilon}_{i}^{-}\right) $. To see how things
work, note that the general form of the wave functions in the full
Hilbert space is generated, in addition to (\ref{e4}), by the
extra monomials
\begin{equation}
\prod_{i\geq 1}^{r_{\mathrm{f}}}\mathbf{\upsilon
}_{k_{i}}^{\dagger
}|s\rangle ,\qquad \prod_{i\geq 1}^{r_{\mathrm{b}}}\mathbf{\Psi }%
_{k_{i}}^{\dagger }|0\rangle
\end{equation}
where $r_{\mathrm{f}}$ and $r_{\mathrm{b}}$ are positive integers,
the vacuum $|s\rangle $ will be specified later. To avoid
confusion, let us rewrite the supersymmetric quantum constraints
(\ref{16f}-\ref{16fg}) into the equivalent forms
\begin{eqnarray}
&&\langle \mathcal{F}|\mathcal{J}_{\Psi }^{0}|\mathcal{F}\rangle
=Nk_{1}
\label{t1} \\
&&\langle \mathcal{F}|\mathcal{J}_{(\mathrm{{b})}}^{K}T_{mn}^{K}|\mathcal{F}%
\rangle =0  \label{t2} \\
&&\langle \mathcal{F}|\mathcal{J}_{\upsilon
}^{0}|\mathcal{F}\rangle =Nk_{2}
\label{t3} \\
&&\langle \mathcal{F}|\mathcal{J}_{(\mathrm{{f})}}^{K}T_{mn}^{K}|\mathcal{F}%
\rangle =0  \label{t4} \\
&&\langle \mathcal{F}|\mathcal{F}^{K}T_{mn}^{K}|\mathcal{F}\rangle
=0 \label{t5}
\end{eqnarray}
and consider the solution of each of them.

\textit{Solving equations} (\ref{t1}-\ref{t2}): The first equation
tells us that the wave function $|\mathcal{F}\rangle $
should have a monomial of type $\prod_{i\geq 1}^{k_{1}}\mathbf{\Psi }%
_{k_{i}}^{\dagger }$. However the second requires that
$|\mathcal{F}\rangle $ should be $SU(N)$ invariant. The
lowest-energy solution $|\mathcal{F}_{\mathrm{b}}^{0}\rangle $
fulfilling these two requirements is that given in (\ref{9})
\begin{equation}
|\mathcal{F}_{\mathrm{b}}^{0}\rangle =\mathcal{N}_{1}\left[
\varepsilon ^{\alpha _{1}\cdots \alpha _{N}}\;\mathbf{\Psi
}_{\alpha _{1}}^{\dagger }\left( \mathbf{\Psi }^{\dagger }\cdot
\mathbf{A}^{\dagger }\right) _{\alpha _{2}}\cdots \left(
\mathbf{\Psi }^{\dagger }\cdot \mathbf{A}^{\dagger N}\right)
_{\alpha _{N}}\right] ^{k_{1}}\;|0\rangle   \label{fb}
\end{equation}
where the index $\mathrm{b}$ refers to the bosonic part
of the full state $|\mathcal{F}%
^{0}\rangle =|\mathcal{F}_{\mathrm{b}}^{0}\rangle \otimes
|\mathcal{F}_{ \mathrm{f}}^{0}\rangle $ and the upper index $0$ to
the lowest energy solution. By using (\ref{142}) and (\ref{fff}),
one can see that the bosonic contribution to the filling factor
$\nu _{\mathrm{b}}$ of the state $|
\mathcal{F}_{\mathrm{b}}^{0}\rangle $ coincides with that given by
relation ( \ref{bff1}). This is exactly the expected value with
the usual Polychronakos quantum shifting $k_{1}$ to $k_{1}+1$.
Here there is no constraint of the (odd) positive integer $k_{1}$
and so the flux number $N_{\phi }^{(\mathrm{{b })}}$ may have any
value. This establishes the features of the U phase of proposition
{\bf(a)}.

\textit{Solving equations} (\ref{t3}-\ref{t4}): Note that the $N$
superpartners $\left( \mathbf{\upsilon } _{n}^{\dagger
},\mathbf{\upsilon }_{n}^{-}\right) $ of the Polychronakos $
\left( \mathbf{\Psi }_{i}^{\dagger },\mathbf{\Psi }_{i}^{-}\right)
$ operators satisfy a $N$-dimensional Clifford algebra
\begin{eqnarray}
&&\left\{ \mathbf{\upsilon }_{m}^{\dagger },\mathbf{\upsilon }
_{n}^{-}\right\} =-i\delta _{mn}  \notag \\
&&\left\{ \mathbf{\upsilon }_{m}^{\dagger },\mathbf{\upsilon
}_{n}^{\dagger }\right\} =\left\{ \mathbf{\upsilon
}_{m}^{-},\mathbf{\upsilon }
_{n}^{-}\right\} =0  \label{16k} \\
&&\left[ \mathcal{J}_{\upsilon }^{0},\mathbf{\upsilon }_{n}^{\pm
}\right] =\pm \mathbf{\upsilon }_{n}^{\pm }.  \notag
\end{eqnarray}
In these relations
\begin{equation}
\mathcal{J}_{\upsilon }^{0}=\sum_{n}i\mathbf{\upsilon
}_{n}^{\dagger } \mathbf{\upsilon }_{n}^{-}
\end{equation}
is the fermionic part of the $U\left( 1\right) $ current of
(\ref{16d}-\ref {16e}). It should be compared with the bosonic
term
\begin{equation}
\mathcal{J}_{\mathbf{\Psi }}^{0}=\sum_{n=1}^{N}\mathbf{\Psi
}_{n}^{\dagger } \mathbf{\Psi }_{n}^{-}
\end{equation}
encountered in section 2. From (\ref{16k}), we learn that
$\mathbf{\upsilon } _{n}^{\pm }$ can be realized as $2^{\left[
\frac{N}{2}\right] }\times 2^{ \left[ \frac{N}{2}\right] }$
matrices acting on $2^{\left[ \frac{N}{2}\right] }$ dimensional
spinors $|s_{\gamma }\rangle $
\begin{eqnarray}
&&\mathbf{\upsilon }_{n}^{-}|s_{\gamma }\rangle =0 \\
&&\mathcal{J}_{\upsilon }^{0}|s_{\gamma }\rangle =0 \\
&&\mathbf{\upsilon }^{\dagger }|s_{\gamma }\rangle =|1,s_{\gamma
}\rangle .
\end{eqnarray}
The vacuum state $|S\rangle $ of the $\mathbf{C}_{kl}^{\pm }$
Clifford algebra can be realized in terms of the above $|s_{\gamma
}\rangle $ by taking tensor products
\begin{equation}
|S\rangle =\prod_{i=1}^{N}|s_{\gamma _{i}}\rangle .
\end{equation}
Since in the algebras (\ref{16j},\ref{16k}) both sets of $\mathbf{C}%
_{kl}^{\dagger }$ and $\mathbf{\upsilon }_{i}^{\dagger }$
operators are nilpotent
\begin{equation}
\left( \mathbf{C}_{kl}^{\dagger }\right) ^{2}=0,\qquad \left( \mathbf{%
\upsilon }_{i}^{\dagger }\right) ^{2}=0
\end{equation}
one has extra constraints, which should be taken into account
in building up the fermionic contribution $|\mathcal{F}_{\mathrm{f}%
}^{0}\rangle $. Due to the nilpotency property (\ref{c0}) as well
as
\begin{equation}
\mathbf{\upsilon }_{i}^{\dagger }\prod_{j=1}^{N}\mathbf{\upsilon }%
_{j}^{\dagger }=0 \label{106}
\end{equation}
$|\mathcal{F}_{\mathrm{f}}^{0}\rangle $ should have no more than
$N^{2}$ creation operators $\mathbf{C}_{kl}^{\dagger }$ and no
more than $N$ operators $\mathbf{\upsilon }_{i}^{\dagger }$. This
feature means a strong constraint  in constructing the solution of
(\ref{t1}-\ref{t5}). It should be linked with the D phase of the
SMM, i.e. proposition {\bf(b)}. With these tools, it is not
difficult to see that the low-energy gauge invariant solution
satisfying (\ref{t3}-\ref{t4}) is
\begin{equation}
|\mathcal{F}_{\mathrm{f}}^{0}\rangle =\mathcal{N}_{2}\left[
\varepsilon ^{i_{1}\cdots i_{N}}\mathbf{\upsilon
}_{i_{1}}^{\dagger }\left( \mathbf{ \upsilon }^{\dagger }\cdot
\mathbf{C}^{\dagger }\right) _{i_{2}}\cdots \left(
\mathbf{\upsilon }^{\dagger }\cdot \mathbf{C}^{\dagger N}\right)
_{i_{N}}\right] ^{k_{2}}|S\rangle \label{ff}
\end{equation}
where the involved products are
\begin{equation}
\mathbf{\upsilon }^{\dagger }\cdot \mathbf{C}_{i}^{\dagger
}=\sum_{j}\mathbf{ \upsilon }_{j}^{\dagger }\cdot
\mathbf{C}_{ji}^{\dagger },\qquad \mathbf{\upsilon }^{\dagger
}\cdot\mathbf{C}_{i}^{\dagger 2}=\sum_{j,k}\mathbf{\upsilon }
_{k}^{\dagger }\cdot \mathbf{C}_{kj}^{\dagger
}\mathbf{C}{_{ji}^{\dagger}}
\end{equation}
and so on. In this relation the charge $k_{2}$ seems to play a
role completely equivalent to the bosonic charge $k_{1}$. However
this is not true because of the nilpotency relations
(\ref{c0},\ref{106}), which require that $k_{2}$ may take only two
values
\begin{equation}
k_{2}=0,1.  \label{k2}
\end{equation}
This restriction is a manifestation of the Pauli exclusion
principle for fermions. But it has several consequences, which can
be listed as follows.

{\bf{(i)}} The flux number $N_{\phi }^{(\mathrm{{f})}}=\left(
1-k_{2}\right) N$ is bounded. This shows that the D phase has a
 %\newline $~\,~\,~\,~$ $~~~$
filling factor $\nu _{\mathrm{f}}$ a priori given by (\ref{fff1}).

{\bf{(ii)}} We distinguish two different solutions:

\underline{Case $k_{2}=0$}: Here the lowest energy configuration
is very special since it corresponds to the vacuum $|S\rangle $,
which is a wave function without $\mathbf{\upsilon}_{i}^{\dagger}$
and $\mathbf{C}_{ij}{^{\dagger}}$ dependence. Its vacuum energy is
given by the classical null value shifted by the negative quantum
correction
\begin{equation}
-\frac{\omega N^{2}}{2}.
\end{equation}
This is a remarkable feature, namely world line fermions induce a
negative quantum flux number
\begin{equation}
\left( k_{2}-1\right) N=-N_{\phi }^{(\mathrm{{f})}}.
\end{equation}
Now by defining $N_{\phi }^{(\mathrm{{f})}}$ as the opposite of
this number
\begin{equation}
N_{\phi }^{(\mathrm{{f})}}=\left( 1-k_{2}\right) N=1
\end{equation}
we obtain an integer filling factor for the D phase. On the other
hand, following (\ref{fo},\ref{fa}) we have
\begin{equation}
\label{totfln}
 N_{\phi }=k_{1}N+\left( N-N\right) =k_{1}N
\end{equation}
where two quantum corrections cancel each other exactly. Then the
total filling factor $\nu_{\rm tot} $ of the ground state
$|\mathcal{F}^{0}\rangle$ is given by
\begin{equation}
\nu_{\rm tot} =\frac{1}{k_{1}}=\frac{1}{k}.  \label{nu}
\end{equation}
Using the relation~(\ref{totfln}), one can see that (\ref{nu}) can
also be rewritten as (\ref{fa}). This establishes the point
{\bf(c)} of our proposition for $k_{2}=0$.

\underline{Case $k_{2}=1$}: In this case, the ground state
$|\mathcal{F}^{0}\rangle $ of the super fractional QH system
depends on the operators $\mathbf{\upsilon } _{i}^{\dagger }$ and
$\mathbf{C}_{kl}^{\dagger }$. Therefore the resulting state can be
written as
%\begin{equation}
\begin{eqnarray}
|\mathcal{F}^{0}\rangle &=&\mathcal{N}\left[ \varepsilon ^{\alpha
_{1}\cdots \alpha _{N}}\Psi _{\alpha _{1}}^{\dagger }\left( \Psi
^{\dagger }\cdot \mathbf{A}^{\dagger }\right) _{\alpha _{2}}\cdots
\left( \Psi ^{\dagger }\cdot \mathbf{A}^{\dagger N}\right)
_{\alpha _{N}}\right] ^{k_{1}}\; \notag \\
&&\times \; \varepsilon ^{i_{1}\cdots i_{N}}\mathbf{\upsilon }
_{i_{1}}^{\dagger }\left( \mathbf{\upsilon }^{\dagger }\cdot
\mathbf{C} ^{\dagger }\right) _{i_{2}}\cdots \left(
\mathbf{\upsilon }^{\dagger }\cdot \mathbf{C}^{\dagger N}\right)
_{i_{N}}\; |0\rangle \otimes |S\rangle.  \label{s1}
\end{eqnarray}
The total number $M$ of $\upsilon _{i}^{\dagger}$ and
$\mathbf{C}_{ij}^{\dagger} $ is equal to
\begin{equation}
M=\frac{N\left( N+3\right) }{2}.
\end{equation}
 From the world line SUSY view,
this state behaves as a bosonic state if $M$ is an even integer
number and as a fermion if $M$ is odd. Now using (\ref{fa}) and
the energies
\begin{equation}
\begin{array}{l}
E_{\mathrm{b}}=\frac{\omega }{2}\left[ k_{1}N\left( N+1\right)
+N^{2}\right]
\\
E_{\mathrm{f}}=\frac{\omega }{2}\left[ N\left( N+1\right) -N^{2}\right] =%
\frac{\omega }{2}N
\end{array}%
\end{equation}
we can show that the total filling factor corresponding to
(\ref{s1}) is
\begin{equation}
\nu_{\rm tot} =\frac{1}{k_{1}+1}=\frac{1}{k}.
\end{equation}
Here also the quantum corrections coming from bosons and fermions
cancel exactly due to the world line SUSY. This tells us that
Polychronakos effects cancel exactly in supersymmetric FQH
systems.

%%%%%%%%%%%%%%%%%%%%%%%%%%%%%%%%%%%%%%%%%%%%%%%%%%
\subsection{Superfield matrix fractional QH model}
%%%%%%%%%%%%%%%%%%%%%%%%%%%%%%%%%%%%%%%%%%%%%%%%%%

Our aim here is to show that, in general, it is more convenient to
adopt the superspace method in the study of our SMM. This is a
manifestly supersymmetric formulation which allows a deeper
insight into the general properties of the fractional QH system
without entering into details about bosonic and fermionic
contributions. It allows, amongst others, a derivation of direct
results extending of the non-commutative Susskind theory for the
Laughlin fluid \cite{susskind}. Here we will mainly focus on
building the superfield extension of the SP model and make a link
with the previous component field analysis.

%%%%%%%%%%%%%%%%%%%%%%%%%%%%%%%%%%%%%%%%%%%%%%%%%%
\subsubsection{$\mathcal{N}=1$ worldline SUSY}
%%%%%%%%%%%%%%%%%%%%%%%%%%%%%%%%%%%%%%%%%%%%%%%%%

The SMM of the fractional QH system, involving
$N_{\mathrm{e}}+N_{\widetilde{\mathrm{e}}}$ particles and
superpartners, has a supersymmetric conserved charge $Q$. This is
related to the usual conserved Hamiltonian $\mathbf{H}_{\rm susy}$
of the fractional QH system by
\begin{equation}
Q^{2}=\mathbf{H}_{\rm susy}.  \label{17}
\end{equation}
This defines a 1D supersymmetric algebra and has irreducible
linear representations $\mathcal{R}_{s}$, which are completely
characterized by a superspin quantum number $\it{s}$, with
$2\it{s}$ integer. In supersymmetric 1D field formulation, a
convenient way to realize  $ \mathcal{R}_{s}$ is to work in the
superspace $\left( t,\zeta \right) $ and use the supersymmetric
covariant derivative $D$
\begin{equation}
DQ=-QD
\end{equation}
satisfying the superalgebra
\begin{equation}
D^{2}=E=i\frac{\partial }{\partial t}.  \label{18}
\end{equation}
In the $\left( t,\zeta \right) $ superspace, $D$ and $E$
generators are related by
\begin{equation}
D=\frac{\partial }{\partial \zeta }+\zeta E=\frac{\partial
}{\partial \zeta } +i\zeta \frac{\partial }{\partial t}
\label{19}
\end{equation}
and $\mathcal{R}_{s}$ are described by 1D superfields $\Phi _{s}$.
These can be expanded as
\begin{equation}
\Phi _{s}\left( t,\zeta \right) =\varphi _{s}\left( t\right)
+\zeta \xi _{s+ \frac{1}{2}}\left( t\right) .  \label{20}
\end{equation}
The scale dimensions of the component fields and their functionals
may be directly obtained by using the scales of $\zeta $ and $t$,
such as $\zeta $, $t$, $D$\ and $\partial _{t}$ carry scale values
$-1/2$, $-1$, $1/2$ and $1$, respectively. For the special example
where  $s=-\frac{1}{2}$, which is the case when the component
field $\xi _{s+\frac{1}{2}}\left( t\right) $ is a 1D scalar, the
SE of the term $\xi ^{1}\partial _{t}\xi ^{2}$ is obtained in
steps, by computing first
\begin{eqnarray}
D\Phi _{s} &=&\xi _{s+\frac{1}{2}}+ i\zeta \partial _{t}\varphi
_{s} \notag
\\
\partial _{t}\Phi _{s} &=&\partial _{t}\varphi _{s}+\zeta \partial _{t}
\xi_{s+\frac{1}{2}} \label{par}
\end{eqnarray}
and by looking second for the appropriate combination that
contains $\xi ^{1}\partial _{t}\xi ^{2}$. Then the SE of $\xi
^{1}\partial _{t}\xi ^{2}$ is
\begin{equation}
\int d\zeta \;D\Phi ^{1}\partial _{t}\Phi ^{2}=\xi ^{1}\partial
_{t}\xi ^{2}+\partial _{t}\varphi ^{1}\partial _{t}\varphi ^{2}.
\label{21}
\end{equation}
In similar fashion we can compute others supersymmetric invariants
and in particular
\begin{equation}
\int d\zeta \Phi D\Phi =\xi \xi +\varphi \partial _{t}\varphi .
\label{22}
\end{equation}
In the presence of the gauge symmetries, $D$ and $\partial _{t}$
should be replaced by the gauge covariant derivatives
\begin{eqnarray}
\mathcal{D} &=&D-i\mathcal{A}=\left( \partial _{\zeta }-i\alpha
\right)
+i\zeta \left( \partial _{t}-i\gamma \right)  \notag \\
\mathcal{\nabla }_t&=&\mathcal{\nabla } =\partial
_{t}-i\mathcal{C} =\partial _{t}-iA + i\zeta \rho   \label{23}
\end{eqnarray}
where $\mathcal{A}$ and $\mathcal{C}$ are superfields related to
each other by the constraint equations
\begin{eqnarray}
\left\{ \mathcal{D},\mathcal{D}\right\} &=&2\mathcal{D}^{2}=2i\mathcal{%
\nabla }  \notag \\
\left[ \mathcal{D},\mathcal{\nabla }\right] &=&0.  \label{24}
\end{eqnarray}
These equations imply, in general, that $\mathcal{A}$ and
$\mathcal{C}$ are not really independent since they are connected
by
\begin{equation}
D\mathcal{A}+\frac{i}{2}\left\{ \mathcal{A},\mathcal{A}\right\}
=\mathcal{C}. \label{acting}
\end{equation}
Also we recall the useful property
\begin{equation}
\int d\zeta \Phi =\mathcal{D}\Phi |_{\zeta =0}  \label{25}
\end{equation}
for explicit computations. Having given the main superspace tools,
now we turn to study the degrees of freedom, which we will use to
build the supersymmetric generalization of the SP theory.

%%%%%%%%%%%%%%%%%%%%%%%%%%%%%%%%%%%%%%%%%%%
\subsubsection{Degrees of freedom}
%%%%%%%%%%%%%%%%%%%%%%%%%%%%%%%%%%%%%%%%%%%

In addition to the superspin described above, the matrix model for
the fractional QH system involves extra quantum numbers associated
with the symmetry group $G=SO\left( 2\right) \times U\left(
N\right) $. Here $SO\left( 2\right) $ is the symmetry group of the
$\mathbb{R}^{2}$ plane and $U\left( N\right) $ the usual gauge
group of the SP model. We need the superfields that transform
under the representations $ \left( \mathbf{r},\mathrm{R}\right) $
of $SO\left( 2\right) \times U\left( N\right) $ and have the
$\zeta $-expansions
\begin{equation}
\Phi _{\left( \mathbf{r},\mathrm{R}\right) }=\varphi _{\left(
\mathbf{r}, \mathrm{R}\right) }+\zeta \xi _{\left(
\mathbf{r},\mathrm{R}\right) }. \label{26}
\end{equation}
For simplicity, we have dropped the superspin index.
the superfield contents of our SMM are\\

\begin{equation}
\begin{tabular}{|l|l|l|l|l|l|l|}
\hline Fields & $\chi ^{a}$ & $\Phi $& $\mathcal{A}$ &
$\mathcal{C}$& $\Gamma $ \\
\hline Superspin $s$ & $-1/2$ & $-1/2$ & $1/2$ & $1$ & $-1/2$
\\ \hline $SO(2)$ repres. $\mathbf{r}$ & $2$ & $1$ & $1$ & $1$ & $1$
\\ \hline
$U(N)$ repres. $\mathrm{R}$ & $\mathbf{N\times
}\overline{\mathbf{N}}$ & $ \mathbf{N}$ & $\mathbf{N\times
}\overline{\mathbf{N}}$ &
$\mathbf{N\times }\overline{\mathbf{N}}$ & $\mathbf{N\times }%
\overline{\mathbf{N}}$
\\ \hline
\end{tabular}
\label{27}
\end{equation}\\

\noindent where $\chi ^{a}$ parameterizes superpositions and
supermomentum of the supersymmetric particles, $\Phi $ is the SE
of the Polychronakos field. $\mathcal{A}$ and $\mathcal{C}$ are
the supersymmetric gauge fields needed to capture the constraint
equations on the super fractional QH system. $\Gamma $ is a
fermionic superfield carrying the supersymmetric constraint
equation (\ref{24}). In terms of the components fields $\left(
\eta ^{a},X^{a}\right) $, $\left( \upsilon ,\Psi \right) $,
$\left( \beta , \mathrm{b}\right) $, $\left( \alpha ,\gamma
\right) $ and $\left( A,\rho \right) $ of $\chi ^{a}$, $\Phi $,
$\Gamma $, $\mathcal{A}$ and $\mathcal{C}$
respectively, the previous table becomes\\

\begin{equation}
\begin{tabular}{|l|l|l|l|l|l|l|}
\hline Fields & $\eta ^{a}=\chi ^{a}|$ & i$X^{a}=\mathcal{D}\chi
^{a}|$ & $\upsilon =\Phi |$ & i$\Psi =\mathcal{D}\Phi |$ & $\alpha
=\mathcal{A}|$ & i$\gamma = \mathcal{DA}|$ \\ \hline Superspin $s$
& $-1/2$ & $0$ & $-1/2$ & $0$ & $1/2$ & $1$ \\ \hline $SO(2)$
repres. $\mathbf{r}$ & $2$ & $2$ & $1$ & $1$ & $1$ & $1$
\\ \hline $U(N)$ repres.
$\mathrm{R}$ & $\mathbf{N\times }\overline{\mathbf{N}}$ &
$\mathbf{N\times } \overline{\mathbf{N}}$ & $\mathbf{N}$ &
$\mathbf{N}$ & $\mathbf{N\times } \overline{\mathbf{N}}$ &
$\mathbf{N\times }\overline{\mathbf{N}}$
\\ \hline
\end{tabular}
\label{28}
\end{equation}\\

\noindent and\\

\begin{equation}
\begin{tabular}{|l|l|l|l|l|}
\hline Fields & $A=\mathcal{C}|$ & i$\rho =\mathcal{D}\Gamma |$ &
$ \QTR{sl}
{\beta }=%
{\Gamma }|$ & i$\mathrm{b}=\mathcal{D}\Gamma |$ \\ \hline
Superspin $s$ & $1$ & $3/2$ & $-1/2$ & $0$ \\ \hline $SO(2)$
repres. $\ \mathbf{r}$ & $1$ & $1$ & $1$ & $1$ \\ \hline $U(N)$
repres. $\mathrm{R}$ & $\mathbf{N\times }\overline{\mathbf{N}}$ &
$ \mathbf{N\times }\overline{\mathbf{N}}$ & $\mathbf{N\times
}\overline{ \mathbf{N}}$ & $\mathbf{N\times
}\overline{\mathbf{N}}$ \\ \hline
\end{tabular}
\label{29}
\end{equation}\\

\noindent where $\chi ^{a}|$ stands for $\chi ^{a}|_{\zeta =0}$
and so on.

%%%%%%%%%%%%%%%%%%%%%%%%%%%%%%%%%%%%%%%%%%%%%%%%%%%%%%%%%%%%%%%
\subsubsection{Superfield constraints}
%%%%%%%%%%%%%%%%%%%%%%%%%%%%%%%%%%%%%%%%%%%%%%%%%%%%%%%%%%%%%%%

One way to construct the superfield constraint equations is to
start from the bosonic constraints and use the ST of the component
fields
\begin{equation}
\begin{array}{l}
\delta _{\epsilon }{\eta }^{a} =\epsilon X^{a},\qquad \delta
_{\epsilon
}X^{a}=-i\epsilon \partial _{t}{\eta }^{a}  \\
\delta _{\epsilon }\upsilon =\epsilon \Psi ,\qquad \delta
_{\epsilon }\Psi
=-i\epsilon \partial _{t}\upsilon   \label{290} \\
\delta _{\epsilon }\zeta =\epsilon
\end{array}
\end{equation}
together with analogous relations for the remaining fields.  In
doing so, one can already derive partial information on the
supersymmetrization of the $ U\left( N\right) $ constraint
equations (\ref{3b}), which by the help of ( \ref{16b}) can be
written as
\begin{eqnarray}
&&F_{mn}=i\frac{B}{2}\varepsilon _{ab}\left[ {\eta
}^{a},X^{b}\right] _{nm}+\upsilon _{n}^{\ast }\Psi _{m}=\zeta
B\theta \delta _{nm}  \label{29a}
\\
&&J_{mn}=i\frac{B}{2}\varepsilon _{ab}\epsilon \left(\left[
X^{a},X^{b}\right] _{nm}+i\left\{ {\ \eta }^{a},\partial _{t}{\eta
}^{b}\right\} _{nm}\right)+\epsilon \Psi _{n}^{\ast }\Psi
_{m}+i\epsilon \upsilon _{n}^{\ast }\partial _{t}\upsilon
_{m}=\epsilon B\theta \delta _{nm} \label{29b}
\end{eqnarray}
where $\zeta $\ is the superspace odd coordinate. As we have done
before, it is convenient to split these constraints into the
irreducible $SU\left( N\right) $ and $U\left( 1\right) $ parts
\begin{equation}
\begin{array}{l}
\mathcal{F}_{mn}=F_{mn}-\frac{1}{N}\mathcal{F}^{0}\\
 \mathcal{J}%
_{mn}=J_{mn}-\frac{1}{N}\mathcal{J}^{0}  \label{feq}
\end{array}
\end{equation}
where the component field expressions for $\mathcal{F}^{0}$ and
$\mathcal{J} ^{0}$ are given by
\begin{eqnarray}
&&\mathcal{F}^{0}=\sum_{n=1}^{N}\upsilon _{n}^{\ast }\Psi
_{n}+\Psi_{n}\upsilon _{n}^{\ast } =2B\theta N\zeta  \label{30a} \\
&&\mathcal{J}^{0}=\sum_{n=1}^{N} \Psi _{n}^{\ast }\Psi _{n}+\Psi
_{n}\Psi _{n}^{\ast } +i\left( \upsilon _{n}^{\ast }\partial
_{t}\upsilon _{n}-\partial _{t}\upsilon _{n}\upsilon _{n}^{\ast
}\right) =2B\theta N.  \label{30b}
\end{eqnarray}
While (\ref{29b},\ref{30b}) are adequately defined, the relations
(\ref{29a},\ref{30a}) break explicitly the SUSY. This difficulty
is due to a supergauge fixing and may be surmounted by working in
superspace and introducing auxiliary superfields. Let us describe
briefly how the machinery works.

\textbf{(1) Supersymmetric action:} In terms of the superfields of
table (\ref{27}), the simplest supersymmetric action $S \left[
\chi ^{a},\Phi ,\mathcal{C},\mathcal{A},\Gamma \right] $, which is
$ U\left( N\right) $ gauge invariant and generalizes the SP matrix
model action (\ref{3}) is
\begin{eqnarray}
S\left[ {\chi }^{a},\Phi ,\mathcal{C},\mathcal{A},\Gamma \right]
&=&-\frac{B }{2}\int dt\;d\zeta \left\{\mathrm{Tr}\left(
\varepsilon _{ab}{\chi }^{a}\mathcal{ \nabla D}{\chi }^{b}+\omega
_{0}{\chi }^{a}\mathcal{D}{\chi }^{a}\right)
+\theta \mathrm{Tr}\mathcal{A} \right\}  \notag \\
&&+\int dt \; d\zeta  \left\{\sum_{n=1}^{N}i\Phi _{n}^{\ast
}\mathcal{\nabla D}\Phi _{n}+\mathrm{Tr}\left[ \Gamma \left(
2D\mathcal{A} -i\left\{ \mathcal{A},\mathcal{A}\right\}
-2\mathcal{C}\right) \right] \right\}\label{act2}
\end{eqnarray}
where the first and the fourth terms are, respectively, the
supersymmetrization of the Susskind and Polychronakos terms. The
second term is a confining superpotential and $\Gamma $ is an
auxiliary superfield carrying (\ref{24}). The covariant
derivatives $\mathcal{D}$ and $\mathcal{ \nabla }$ act on
superfields $\Upsilon $ in the $U\left( N\right) $ adjoint as
\begin{eqnarray}
\mathcal{D}{\Upsilon } &=&\partial _{\zeta }\Upsilon -i\left[
\mathcal{A},{\Upsilon }
\right\}   \notag \\
\mathcal{\nabla }{\Upsilon } &=&\partial _{t}\Upsilon -i \left[
\mathcal{C},{\Upsilon } \right]   \label{31}
\end{eqnarray}%
and on superfields in the fundamentals
\begin{eqnarray}
\mathcal{D}\Phi  &=&\partial _{\zeta }\Phi -i\mathcal{A}\Phi   \notag \\
\mathcal{\nabla }\Phi  &=&\partial _{t}\Phi -i\mathcal{C}\Phi.
\label{32}
\end{eqnarray}%
Before expanding the superfield action (\ref{act2}) in terms of
the component fields, it is convenient to rewrite the above
superfield action by using the complex superfield
\begin{equation}
\chi =\frac{1}{\sqrt{B}}\left( {\chi }^{1}+i{\chi }^{2}\right)
={\eta } +\zeta Z  \label{32a}
\end{equation}
where the component fields are given by
\begin{equation}
\eta =\frac{1}{\sqrt{B}}\left( \eta ^{1}+i\eta ^{2}\right)
\end{equation}
and $Z$ is given in subsection 2.1. In terms of this complex
superfield, (\ref {act2}) takes the form
\begin{eqnarray}
S &=&\int dt\;d\zeta \;\mathrm{Tr}\left[ \frac{i}{4}\left(
\overline{\chi } \mathcal{\nabla D}{\chi -}\overline{\left(
\mathcal{\nabla D}{\chi }\right) } {\chi }\right) -\frac{\omega
_{0}}{4}\left( \overline{{\chi }}\mathcal{D}{ \chi
+}\overline{\left( \mathcal{D}{\chi }\right) }{\chi }\right)
\right] +
\notag \\
&&\int dt\;d\zeta \left\{\sum_{n=1}^{N}\frac{i}{2}\left(
\overline{\Phi }_{n} \mathcal{\nabla
D}\Phi_{n}-\overline{\mathcal{\nabla D}\Phi }_{n}\Phi _{n}\right)
+\mathrm{Tr}\left[ {\Gamma }\left( 2D \mathcal{A}-i\left\{
\mathcal{A},\mathcal{A}\right\} -2\mathcal{C}\right)\right]
\right\} . \label{32b}
\end{eqnarray}
It can be further simplified by adding total superderivatives, a
property which we will often use in the coming computations. From
this action, one can compute the various superfield equations of
motion. For dynamical $\chi $ and $\Phi $ superfields, we have
\begin{equation}
\begin{array}{l}
\mathcal{\nabla D}{\chi }+i\omega _{0}\mathcal{D}{\chi } =0  \notag \\
\mathcal{\nabla D}\Phi  =0.  \label{33}
\end{array}
\end{equation}
Using the $\zeta $ expansion (\ref{32a}), we can also derive the
component field equations of motion for $X$, $\eta $, $\upsilon $
and\ $\Psi $. By using (\ref{32a}) as
\begin{eqnarray}
&&\mathcal{\nabla }Z+i\omega _{0}Z=0  \label{34a} \\
&&\mathcal{\nabla }^{2}\eta +i\omega _{0}\mathcal{\nabla }\eta =0
\label{34b} \\
&&\mathcal{\nabla }\Psi =0  \label{34c} \\
&&\mathcal{\nabla }^{2}\upsilon =0.  \label{34d}
\end{eqnarray}%
Note that the superfield action as well as the various equations
of motion are invariant under the $U\left( N\right) $ gauge
symmetry
\begin{equation}
\chi \rightarrow e^{i\Lambda }\chi e^{-i\Lambda },\qquad \Phi
\rightarrow e^{i\Lambda }\Phi   \label{40}
\end{equation}
where
\begin{equation}
\Lambda \left( t\right) =\lambda \left( t\right) +\zeta \kappa
\left( t\right)
\end{equation}
is an arbitrary hermitian matrix gauge superfield parameter of
$U\left( N\right) $.

\textbf{(2) Superconstraints:} To obtain the superfield constraint
equations, we can minimize the action with respect to the
nondynamical superfields $\Gamma ,$ $\mathcal{C}$ and
$\mathcal{A}$. Using the identities
\begin{eqnarray}
&&\mathrm{Tr}\overline{\chi }\left[ \mathcal{\nabla
},\mathcal{D}\chi \right] = \mathrm{Tr}\mathcal{\nabla }\left[
\mathcal{D}\chi ,\overline{\chi }
\right]   \notag \\
&&\mathrm{Tr}\overline{\chi }\left\{ \mathcal{D},\chi \right\}
=\mathrm{Tr}
\mathcal{D}\left\{ \overline{\chi },\chi \right\}  \label{ident} \\
&&\mathrm{Tr}\overline{\chi }\left\{ \mathcal{D},\mathcal{\nabla
}\chi \right\}  =\mathrm{Tr}\mathcal{D}\left\{ \overline{\chi
},\mathcal{\nabla } \chi \right\}   \notag
\end{eqnarray}
and the cyclic property of the trace as well as the statistics of
the superfields, we find the equations of motion for the remaining
superfields $\Gamma , $ $\mathcal{C}$ and $\mathcal{A}$
\begin{eqnarray}
&&2D\mathcal{A}-i\left\{ \mathcal{A},\mathcal{A}\right\}
-2\mathcal{C}=0
\label{35a} \\
&&\frac{1}{2}\left[ \mathcal{D}{\chi }, \overline{\chi }\right]
+\mathcal{D}
\Phi \overline{\Phi }-2\Gamma =0  \label{35b} \\
&&i\frac{B}{2}\varepsilon _{ab}\left[ \chi ^{a},\mathcal{D}\chi
^{b}\right]
+\Phi ^{\ast }\mathcal{D}\Phi -2\Gamma =0  \label{350} \\
&&\frac{1}{2}\left\{ \overline{\chi },\mathcal{\nabla }\chi
\right\} +i\frac{ \omega _{0}}{2}\left\{ \overline{\chi }, \chi
\right\} +\mathcal{\nabla }\Phi
\overline{\Phi }-2\mathcal{D}\Gamma =B\theta   \label{35c} \\
&&-\frac{B}{2}\left( \varepsilon _{ab}\left\{ \chi
^{a},\mathcal{\nabla }{ \chi }^{b}\right\} +\omega _{0}\left\{
\chi ^{a}\chi ^{a}\right\} \right) +i\Phi ^{\ast }\mathcal{\nabla
}\Phi -2\mathcal{D}\Gamma =2B\theta I. \label{351}
\end{eqnarray}
Let us comment the contents of these relations. (\ref{35a})
contains two component field terms. It is a standard super
relation which relates the various degrees of freedom of the
components of the gauge superfields $\mathcal{A}$ and
$\mathcal{C}$. It splits into
\begin{equation}
\begin{array}{l}
\gamma -i\mathcal{\alpha }^{2}-A=0 \\
i\nabla _{t}\alpha =\rho \label{37}%
\end{array}%
\end{equation}%
where now $\nabla _{t}=\mathcal{\nabla }|=\partial _{t}-i\left[
A,~\right] $ for the adjoint matrix fields. Note that in the
special gauge $\alpha=\rho =0$, the gauge field $\gamma $
appearing in the expansion of $\mathcal{A}$ is just the gauge
potential involved in $\mathcal{C}$. (\ref{35b}) gives the
explicit expression of the superfield $\Gamma $ in terms of $\chi
$ and $\Phi $. Its expansion in component fields leads to the
constraints
\begin{eqnarray}
&&i\frac{B}{2}\varepsilon _{ab}\left[ {\eta }^{a},X^{b}\right]
+\upsilon
^{\ast }{\Psi }-2\beta =0 \\
&&-\frac{iB}{2}\varepsilon _{ab}\left( \left[ X^{a},X^{b}\right] +i\left\{ {%
\eta }^{a},\nabla _{t}{\eta }^{b}\right\} \right) -\Psi ^{\ast }{\Psi }%
-i\upsilon ^{\ast }\nabla \upsilon -2ib=0.
\end{eqnarray}%
Upon setting
\begin{equation}
A=0,\qquad 2\beta =\zeta B\theta ,\qquad 2ib=-B\theta \label{set}
\end{equation}%
one discovers exactly (\ref{29a}-\ref{29b}). Furthermore substituting (\ref%
{35b}) into (\ref{351}), we obtain
\begin{equation}
-i\frac{B}{2}\varepsilon _{ab}\left[ \mathcal{D}{\chi
}^{a},\mathcal{D}{\chi
}^{b}\right] -B\varepsilon _{ab}\left\{ {\chi }^{a},\mathcal{\nabla }{\chi }%
^{b}\right\} -\frac{\omega _{0}}{2}\left\{ {\chi }^{a},{\chi }^{a}\right\} -%
\mathcal{D}\Phi ^{\ast }\mathcal{D}\Phi +2i\Phi ^{\ast }\mathcal{\nabla }%
\Phi =\theta BI.
\end{equation}%
The first component field projection of the above superfield
equation is
\begin{equation}
i\frac{B}{2}\varepsilon _{ab}\left[ X^{a},X^{b}\right]
-B\varepsilon _{ab}\left\{ {\eta }^{a},\mathcal{\nabla }{\eta
}^{b}\right\} -\frac{\omega _{0}}{2}\left\{ {\eta }^{a},{\eta
}^{a}\right\} +\Psi ^{\ast }\Psi +2i\upsilon ^{\ast
}\mathcal{\nabla }\upsilon =\theta BI.
\end{equation}%
It reduces to (\ref{29b}) by using the equation of motion for $A$
and setting $A=0$. We now make some simplifications by working in
the remarkable gauge $\alpha =0$.

%%%%%%%%%%%%%%%%%%%%%%%%%%%%%%%%%%%%%%%%%%%%%%%%%%%%%%%%%%%%%%%%%%%%%
\subsubsection{Component field action in gauge\ $\protect\alpha =0$}
%%%%%%%%%%%%%%%%%%%%%%%%%%%%%%%%%%%%%%%%%%%%%%%%%%%%%%%%%%%%%%%%%%%%%

The supersymmetric action can be simplified by fixing a gauge such
as $\alpha=0$. This can be done by expanding the superfields in
terms of the component fields
\begin{equation}
\chi =\eta +\zeta Z,\qquad \Phi =\upsilon +\zeta \Psi,
\label{xset}
\end{equation}
putting them into (\ref{act2}) and using the supersymmetric
algebra $ \mathcal{D}^{2}=i\mathcal{\nabla }$, by taking into
account the statistics of the superfields and the integral
measure.  We find, up to total time derivatives, the component
fields action
\begin{equation}
S\left[ X,\eta ,\Psi ,\upsilon ,A\right] =\int dt\;\mathcal{L}
\label{saction}
\end{equation}%
where the Lagrangian $\mathcal{L}$ is given by
\begin{equation}
\mathcal{L}= \mathrm{Tr}\left[
\frac{i}{2}\overline{Z}\mathcal{\nabla }Z- \frac{1}{2}\left(
\overline{\mathcal{\nabla }\eta }\right) \mathcal{\nabla } \eta
-\frac{\omega _{0}}{2}\left( \overline{Z}Z-i\overline{\eta
}\mathcal{ \nabla }\eta \right) \right] +i\overline{\Psi
}\mathcal{\nabla }\Psi - \overline{\mathcal{\nabla }\upsilon
}\mathcal{\nabla }\upsilon +B\theta \mathrm{Tr}A  \label{38}
\end{equation}%
where $A$ is now the unique auxiliary field. In this case the
gauge covariant derivatives are defined by
\begin{equation}
\begin{array}{l}
\nabla _{t}Z_{nm} =\partial _{t}Z_{nm}-i\left[ A,Z\right] _{nm}  \notag \\
\nabla _{t}\eta _{nm} =\partial _{t}\eta _{nm}-i\left[ A,\eta
\right]_{nm}  \notag \\
\nabla _{t}\Psi _{n} =\partial _{t}\Psi _{n}-iA_{nm}\Psi _{m}
\label{45}
\\
\nabla _{t}\upsilon _{n} =\partial _{t}\upsilon
_{n}-iA_{nm}\upsilon _{m}. \notag
\end{array}
\end{equation}
Using the identities
\begin{equation}
\begin{array}{l}
\mathrm{Tr}\left( \left[ A,X^{a}\right] X^{b}\right)
=\mathrm{Tr}\left( A
\left[ X^{b},X^{a}\right] \right)   \notag \\
\mathrm{Tr}\left( \left[ A,\eta ^{a}\right] \eta ^{b}\right)
=\mathrm{Tr} \left( A\left\{ \eta ^{a},\eta ^{b}\right\} \right)
\label{ident2}
\end{array}
\end{equation}
one can calculate the equation of motion for $A$. We find
\begin{equation}
\frac{1}{2}\left[ Z,\overline{Z}\right] +i\left\{ \overline{\eta
},\mathcal{ \nabla }\eta \right\} +\frac{\omega _{0}}{2}\left\{
\overline{\eta },\eta \right\} +\Psi \Psi ^{\ast
}-2\mathcal{\nabla }\upsilon \upsilon ^{\ast }=B\theta .
\end{equation}
This relation should be compared with
(\ref{16c1},\ref{16c},\ref{16c3}). With this link, one can build
the lowest energy configuration by solving the constraint
equations in a similar fashion to those we have done in subsection
3.2.

%%%%%%%%%%%%%%%%%%%%%%%%%%%%%%%%%
\section{Conclusion}
%%%%%%%%%%%%%%%%%%%%%%%%%%%%%%%%%

We have developed a supersymmetric extension of the
Susskind--Polychronakos matrix model. This is done by making use
of component field and superfield methods. The one-dimensional
$N\times N$ hermitian matrix field $X^{i}$ and its superpartner
$\eta ^{i}$ are involved. Our model has a
supersymmetric U$\left( N\right) $ gauge invariant action (\ref{16B},\ref%
{act2},\ref{32b}) containing the SP model. It is proposed to
describe in this way a class of fractional QH systems. Our results
can be summarized as follows:

{\bf(1)} We have shown that the bosonic $X^{i}$ and fermionic
fields $\eta ^{i}$, which are rotated under world line
supersymmetry, have a gauge invariant vacuum state. This describes
the lowest Landau level ground state with two phases U and D.
{\bf(a)} U is filled by bosonic excitations with no restriction on
the flux number $N_{\phi}=k_{1}N$ since $k_{1}$ can be any
positive (odd)
integer. {\bf(b)} D contains world line fermions with flux number $%
N_{\phi }=k_{2}N$ with $k_{2}=0,1$. Since the filling of U and D
is governed by the statistics of the quantum excitations, it is
interesting to note that the similarity between the present
analysis and Bose-Fermi phases of the quantum approximation for
the ideal gas. In this approach, the U phase of our super matrix
model for the fractional QH droplet can be seen as a Bose-like
condensed state but with a filling constrained by $U\left(
N\right) $ gauge invariance. D is associated with the Fermi
surface of the Fermi gas. Like in U, here also the $U\left(
N\right) $ gauge invariance puts a constraint on the radius of the
Fermi surface.

{\bf(2)} We have constructed the ground state of the
supersymmetric fractional QH system and shown that there is no
Polychronakos shift for the filling factor in the supersymmetric
case, because in our SE, there are two contributions to the
Polychronakos effect. The first one is generated from bosonic
excitations and the second one sums from fermionic excitations.
These contributions are equal but with opposite signs, fermions
generate an anti-Polychronakos effect. This exact cancellation is
due to the world line supersymmetry.

{\bf(3)} We have also given the general relation between the total
energy $E$ of the system and the filling factor $\nu $
\begin{equation}
\nu =\frac{\omega N^{2}}{2E}=\nu _{0}+O\left( \frac{1}{N}\right).
\end{equation}
This is used to study $\frac{1}{N}$\ corrections in fractional QH
systems and in particular allowed us to derive a description of a
Hall system combining bosons and fermions.

Of course still some important questions remain to be answered,
e.g., about the fractional charge and the statistics of the
particles and how to describe them in terms of the proposed model.
Another interesting question is related to the link between our
model and super--Calogero~\cite{group3} and
super--Calogero--Sutherland~\cite{brink} models. We will return to
these issues and related matter in future.

%%%%%%%%%%%%%%%%%%%%%%%%%%%%%%%%%%%%%%%%%%%%%%%%%%%%%%%%%%%%%
\section*{{Acknowledgment}}
%%%%%%%%%%%%%%%%%%%%%%%%%%%%%%%%%%%%%%%%%%%%%%%%%%%%%%%%%%%%%

Three of the authors (SJG, AJ \& EHS) wish to acknowledge the
Stellenbosch Institute for Advanced Study (STIAS, Dir. Bernard
Lategan) workshop on String Theory and Quantum Gravity, for
hospitality in the period of Jan. 23 to  Feb. 22, 2002 during
which the initial discussion of this work began.
 SJG wishes to acknowledge support in part by
the U.S. National Science Foundation under grant number
PHY-0354401. AJ's work is supported partially by Deutsche
Forschungsgemeinschaft within the Schwerpunkt
``Quantum-Hall-Effekt''. EHS thanks D12/25, Protars III
CNRST-Rabat.

\end{document}